\documentclass[a4paper,11pt]{article}
\pdfoutput=1
\usepackage{jheppub}
\usepackage{rotating}
\usepackage{array}
\usepackage{amsmath}
\usepackage[normalem]{ulem}
\usepackage{slashed}
\usepackage{booktabs}
\usepackage[pdftex,table]{xcolor}
\usepackage{units}
\usepackage{xfrac}
\usepackage{mathtools}
\usepackage{empheq}
\usepackage[]{units}
\usepackage{multirow}
\usepackage{amssymb}
\usepackage{url}
\usepackage{comment}
\usepackage{paralist}
\usepackage{xspace}
\usepackage{indentfirst}
\usepackage[symbol]{footmisc}
\usepackage{booktabs}

\usepackage{float}
\usepackage{framed}
\usepackage{graphicx}

\graphicspath{{figures/}}

\usepackage{tikz}

\def\beq{\begin{equation}}
\def\eeq{\end{equation}}
\newcommand{\bea}{\begin{eqnarray}\begin{aligned}}
\newcommand{\eea}{\end{aligned}\end{eqnarray}}
\def\bitem{\begin{itemize}}
\def\eitem{\end{itemize}}

\newcommand{\taujet}{\ensuremath{\tau_{\text{had}}}\xspace}
\newcommand{\pt}{\ensuremath{p_\text{T}}\xspace}
\newcommand{\et}{\ensuremath{E_\text{T}}\xspace}
\newcommand{\trackpt}{\ensuremath{p_\text{T}^{\text{track}}}\xspace}
\newcommand{\tracketa}{\ensuremath{\eta^{\text{track}}}\xspace}
\newcommand{\trackphi}{\ensuremath{\phi^{\text{track}}}\xspace}
\newcommand{\dzero}{\ensuremath{d_0}\xspace}
\newcommand{\zzero}{\ensuremath{z_0}\xspace}
\newcommand{\towerpt}{\ensuremath{E_\text{T}^{\text{tower}}}\xspace}
\newcommand{\towereta}{\ensuremath{\eta^{\text{tower}}}\xspace}
\newcommand{\towerphi}{\ensuremath{\phi^{\text{tower}}}\xspace}
\newcommand{\jetpt}{\ensuremath{p_\text{T}^{\text{jet}}}\xspace}
\newcommand{\jeteta}{\ensuremath{\eta^{\text{jet}}}\xspace}
\newcommand{\jetphi}{\ensuremath{\phi^{\text{jet}}}\xspace}

\newcommand{\trackdeta}{\ensuremath{\eta^{\text{track}} - \eta^{\text{jet}}}\xspace}
\newcommand{\trackdphi}{\ensuremath{\phi^{\text{track}} - \phi^{\text{jet}}}\xspace}
\newcommand{\towerdeta}{\ensuremath{\eta^{\text{tower}} - \eta^{\text{jet}}}\xspace}
\newcommand{\towerdphi}{\ensuremath{\phi^{\text{tower}} - \phi^{\text{jet}}}\xspace}
\newcommand{\fracltrack}{\ensuremath{f_{\text{track}}}}
\newcommand{\rtrack}{\ensuremath{R_{\text{track}}}}
\newcommand{\mtrack}{\ensuremath{m_{\text{track}}}}
\newcommand{\nisotrack}{\ensuremath{N_{\text{track}}^{\text{iso}}}}
\newcommand{\drtrack}{\ensuremath{\Delta R_{\text{track}}}}

\newcommand{\homoenc}{\textit{Homogeneous Encoder}}
\newcommand{\hetnode}{\textit{Heterogeneous Node Encoder}}
\newcommand{\hetedge}{\textit{Heterogeneous Node and Edge Encoder}}
\newcommand{\recrenc}{\textit{Recurrent Encoder}}
\newcommand{\attnenc}{\textit{Attention Encoder}}

\definecolor{darkpurple}{rgb}{0.5, 0.2, 0.8}
\definecolor{darkblue}{rgb}{0.0, 0.0, 0.8}
\definecolor{darkgreen}{rgb}{0.0, 0.4, 0.0}
\definecolor{darkred}{rgb}{0.5, 0.0, 0.0}
\usepackage{hyperref}
\hypersetup{
    linktocpage,
     colorlinks,
     citecolor=darkgreen,
     linkcolor= darkgreen,
     urlcolor=darkgreen
}
\usepackage{lineno}
\abstract{
We present a new algorithm that identifies reconstructed jets originating from hadronic decays of tau leptons against those from quarks or gluons. No tau lepton reconstruction algorithm is used. Instead, the algorithm represents jets as heterogeneous graphs with tracks and energy clusters as nodes and trains a Graph Neural Network to identify tau jets from other jets. Different attributed graph representations and different GNN architectures are explored. We propose to use differential track and energy cluster information as node features and a heterogeneous sequentially-biased encoding for the inputs to final graph-level classification. 
}


\begin{document}

\title{Heterogeneous Graph Neural Network for Identifying Hadronically Decayed Tau Leptons at the High Luminosity LHC}

\author[a]{Andris Huang,}
\author[b]{Xiangyang Ju,}
\author[a]{Jacob Lyons,}
\author[b]{Daniel Murnane,}
\author[c]{Mariel Pettee,}
\author[d]{and Landon Reed}

\affiliation[a]{Physics Department, University of California, Berkeley, CA 94720, USA}
\affiliation[b]{Scientific Data Division, Lawrence Berkeley National Laboratory, Berkeley, CA 94720, USA}
\affiliation[c]{Physics Division, Lawrence Berkeley National Laboratory, Berkeley, CA 94720, USA}
\affiliation[d]{Physics Department, University of Minnesota Duluth, Duluth, MN 55812, USA}

\emailAdd{andrewhz@berkeley.edu}
\emailAdd{xju@lbl.gov}
\emailAdd{jacoblyons98@berkeley.edu}
\emailAdd{dtmurnane@lbl.gov}
\emailAdd{mpettee@lbl.gov}
\emailAdd{reed0658@d.umn.edu}

\maketitle
\flushbottom

%===================================================================
\section{Introduction} \label{sec:intro}
%===================================================================
Tau leptons are important in many physics programs in ATLAS~\cite{ATLAS:2008xda} and CMS~\cite{CMS:2008xjf}. Examples are the measurements of the Higgs boson~\cite{ATLAS:2015xst,ATLAS:2015nvm,ATLAS:2016ifi,ATLAS:2018ynr,ATLAS:2020evk,ATLAS:2022yrq,CMS:2017zyp,CMS:2021sdq,CMS:2021gxc,CMS:2020mpn,CMS:2022kdi,Collaboration:2022mlq} and the search for additional Higgs boson~\cite{ATLAS:2012jag,ATLAS:2017eiz,CMS:2018rmh,CMS:2019kca,CMS:2019pex}. These analyses highly depend on the efficiency and accuracy of the tau reconstruction and identification. To this end, both ATLAS and CMS combined the tracking information reconstructed from hits recorded by their inner detectors with the energy cluster information measured by their calorimeters. ATLAS developed a \textit{Tau Particle Flow} method. Firstly, the kinematics of charged and neural pions are reconstructed, respectively. Then both pieces of information are used as inputs to train a Boost Decision Tree (BDT) to distinguish $\tau$ leptons~\cite{ATLAS:2015boj}. CMS developed a deep neural network, specifically the Convolutional Neural Network (CNN), which takes as inputs the physics-inspired features combined with hidden features learned from energy deposits in the calorimeter and outputs a multi-class score that discriminates $\tau$ leptons against jets, electrons, and muons~\cite{CMS:2022prd}. In addition, ATLAS used a Recurrent Neural Network (RNN) to identify hadronic tau \taujet decays~\cite{ATLAS:2019uhp}. The relational information between tracks and energy clusters explored by ATLAS and CMS is mostly based on physics inspirations, serving as inputs to the \taujet identification algorithms. We propose to use Graph Neural Network to learn the relational information between tracks and towers for the \taujet identification.

Tau lepton decays hadronically 65\% of the time and leptonically 25\% of the time. In the hadronic decay mode, tau leptons primarily decay to one charged pion (70\%) or to three charged pions (21\%), and the majority (68\%) include one or more neutral pions. Therefore, their experimental signature corresponds to a jet with one or three tracks in the detector. The former signature is called \textit{one prong} and the latter \textit{three prongs}. Neutrinos from the hadronic tau lepton decay may be inferred from the missing transfer momentum but cannot be reconstructed. This paper focuses on hadronic tau decays.

Recent years have seen many successful applications of treating proton-proton ($pp$) collision events as graphs and using Graph Neural Networks (GNN) to identify physics objects in particle physics. Ref~\cite{1808887} reviews general GNN applications in particle physics, and Ref~\cite{Duarte:2020ngm} focuses on applications in particle tracking and reconstruction. Graphs can naturally represent collision events at different levels depending on the objective of the task. At a high-level, graphs can represent the collision events with reconstructed physics objects as nodes. At a lower level, graphs can represent individual physics objects with detector-level objects such as tracks and energy clusters/towers as nodes. In both cases, graph edges may or may not exist, and nodes are of different types, making those graphs heterogeneous. Heterogenous graphs can be treated as homogeneous graphs by selecting a subset of common node and edge features at the cost of information loss or by padding node features with zeros at the cost of computations. We represent heterogeneous GNNS that treat different node and edge types differently.

This work is organized as follows. Section~\ref{sec:data} describes the simulated dataset, followed by Section~\ref{sec:method} explaining different graph representations and GNN architectures. Numerical results are shown in Section~\ref{sec:results}. And we discuss our findings in Section~\ref{sec:discuss} and present conclusions in Section~\ref{sec:conclusions}.

\section{Simulation} \label{sec:data}
Proton-proton collisions are simulated with \textsc{Pythia 8.302}~\cite{Sjostrand:2019zhc,Sjostrand:2014zea} at a center-of-mass-energy of $\sqrt s = 13$ TeV. The detector response is simulated by \textsc{Delphes 3}~\cite{deFavereau:2013fsa} with the ATLAS detector configuration, and an average of 200 additional $pp$ collisions emulating the collision density at the High Luminosity LHC are added to each event.  \textsc{Delphes} uses the particle-flow reconstruction algorithms to produce \textit{particle-flow tracks} and \textit{particle-flow towers}, which then serve as inputs to reconstruct jets and missing transverse energy. Jets are reconstructed with the anti-$k_t$~\cite{anti-kt}  algorithm with a radius of 0.4 using \textsc{FastJet 3.3.2}~\cite{Cacciari:2011ma,fastjet}. As no $\tau$ reconstruction algorithm is used, all jets with $\pt > 30$ GeV and $|\eta| < 2.7$ are treated as \taujet candidates.

Genuine \taujet leptons are simulated by the $pp \to \gamma^* \to \tau \tau$ processes, and fake \taujet candidates by jets from the Quantum chromodynamics (QCD) processes. Note that the on-shell $Z$ boson production is excluded to avoid possible biases on the jet kinematic variables. Both are generated by \textsc{Pythia 8.302}, which is also used for parton shower, hadronization, and $\tau$ lepton decay. The A14~\cite{TheATLAScollaboration:2014rfk} set of 
tuned parameters and the \textsc{NNPDF2.3 LO}~\cite{NNPDF:2014otw} set of parton distribution functions with $\alpha_s = 0.13$ are used. All tau leptons were forced to decay hadronically to maximize the generation efficiency. In addition, the invariant mass of two tau leptons ($m_{\tau\tau}^{*}$) is set to be in the range from 60 GeV to 7000 GeV. To populate generated jet \pt spectrum towards high values, the QCD events use a biased phase space sampling which is compensated by a continuously decreasing weight for the event.

We generate 964,000 ditau events and 885,000 QCD events, of which 80\% are used for training, 10\% for validation, and 10\% for testing. A jet is matched to a true $\tau$ lepton if the angular distance $\Delta R$~\footnote{$\Delta R$ is the Euclidean distance in the transverse plane, defined as $\sqrt{\Delta \eta^2 + \Delta \phi^2}$} between the two objects is less than 0.2. We label a jet as a real \taujet if a jet is matched to a true $\tau$ lepton; otherwise, a QCD jet. With that, there are 134,000 true 1-Prong \taujet jets, 75,000 true 3-Prong \taujet jets (total 209,000 true \taujet jets), and 2,800,000 QCD jets in the generated events. Given the imbalances of the true \taujet jets and QCD jets, we assign a larger weight to true tau jets in the loss function so that the total weights of the two classes are close.

\begin{figure}[htb]
    \centering
    \includegraphics[width=0.49\textwidth]{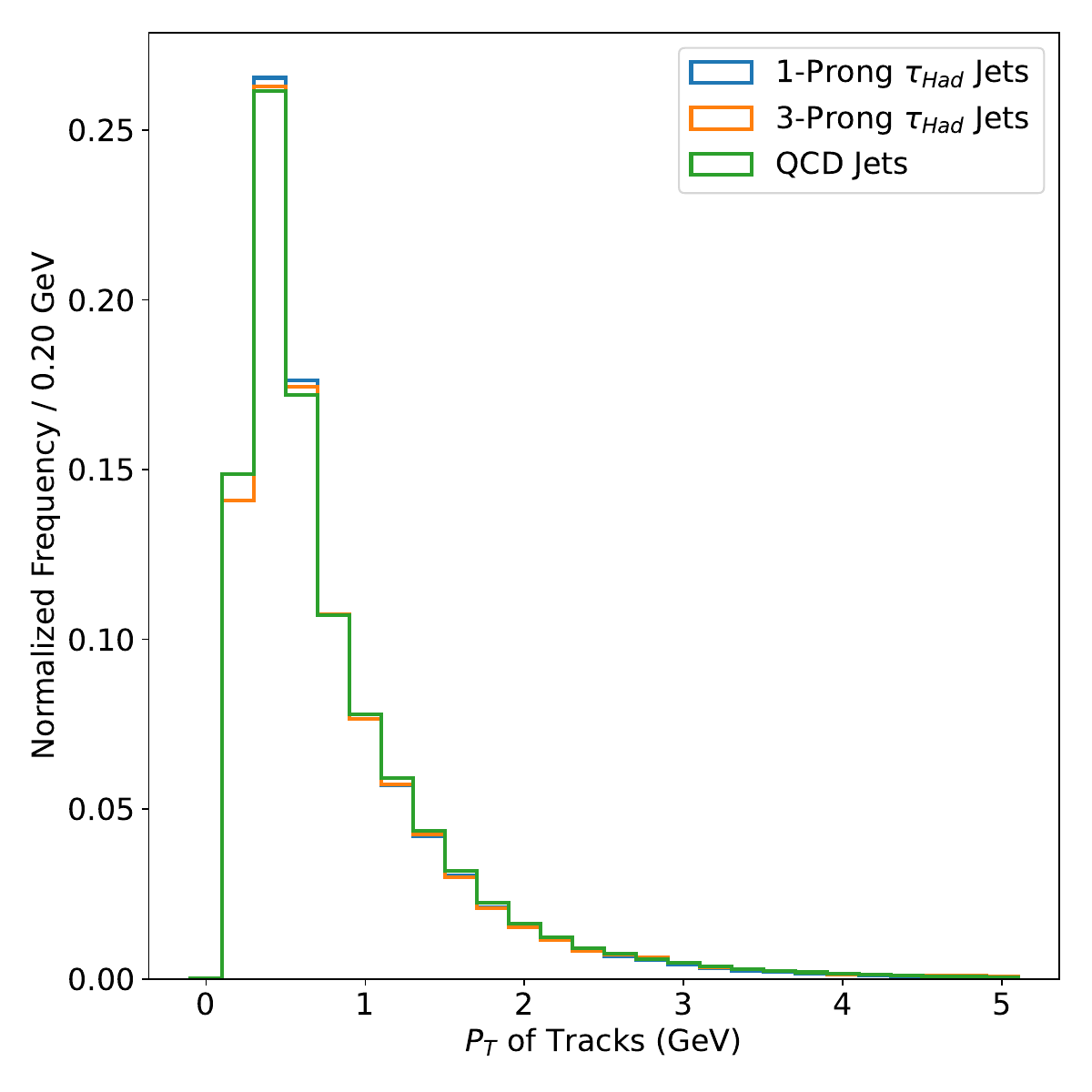}
    \includegraphics[width=0.49\textwidth]{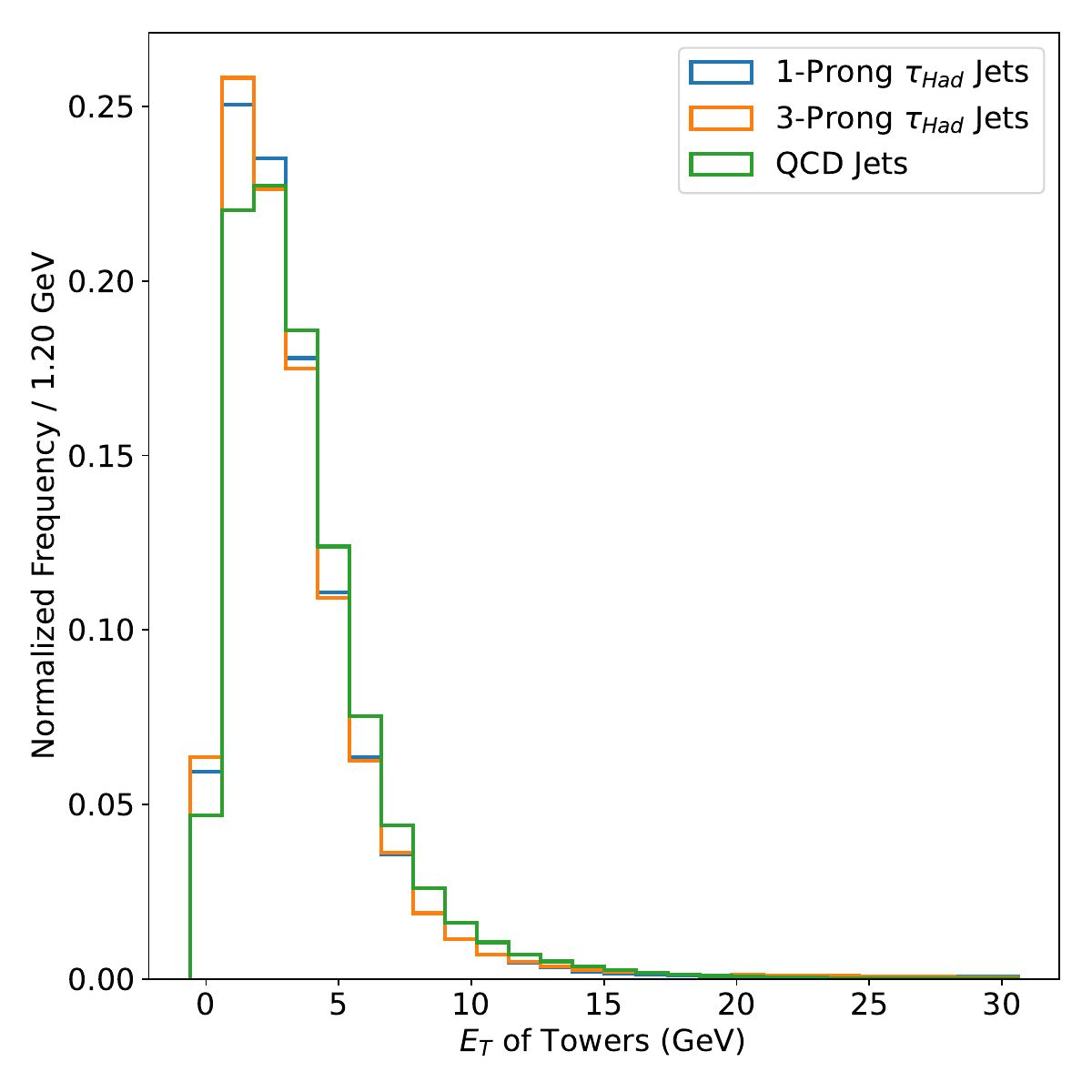} 
    \includegraphics[width=0.49\textwidth]{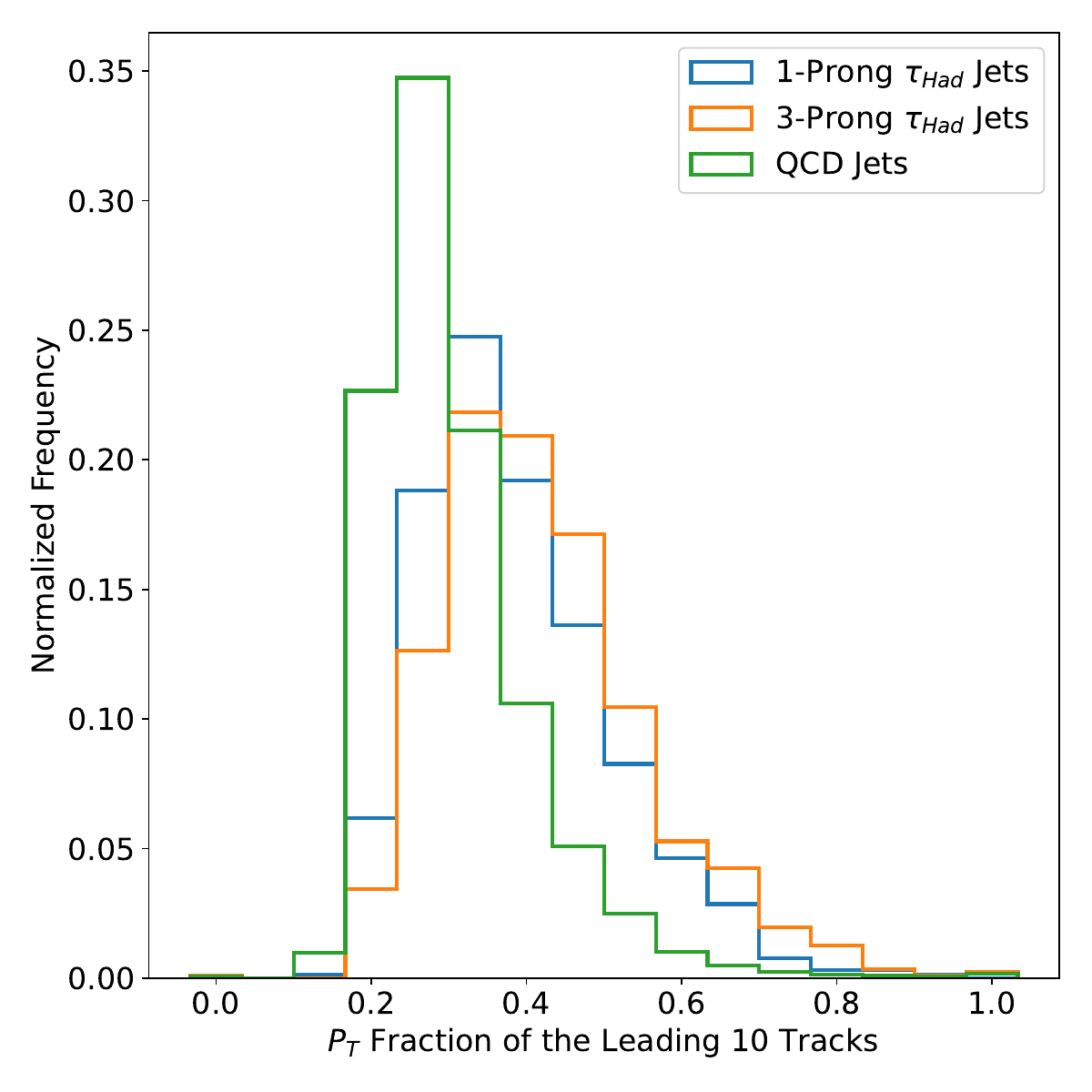}
    \includegraphics[width=0.49\textwidth]{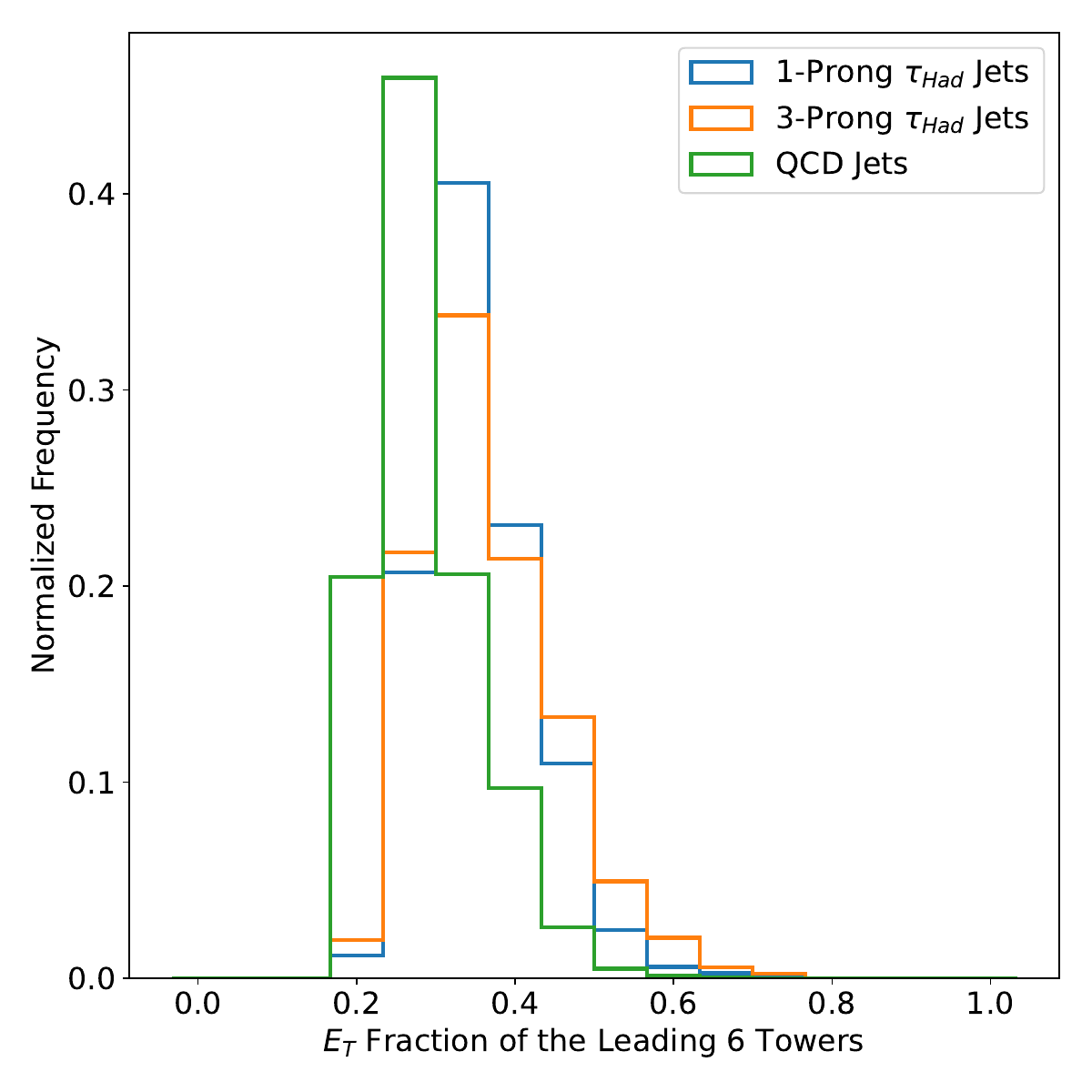}
    \caption{Comparison of the \pt and \et of 1-Prong \taujet jets, 3-Prong \taujet jets, and background (QCD jets). All distributions are normalized to the same area. The fraction of \pt (\et) of the 10 (6) leading tracks (towers) is shown in the bottom rows.}
    \label{fig:pt_et}
\end{figure}

\begin{figure}[htb]
    \centering
    \includegraphics[width=0.31\textwidth]{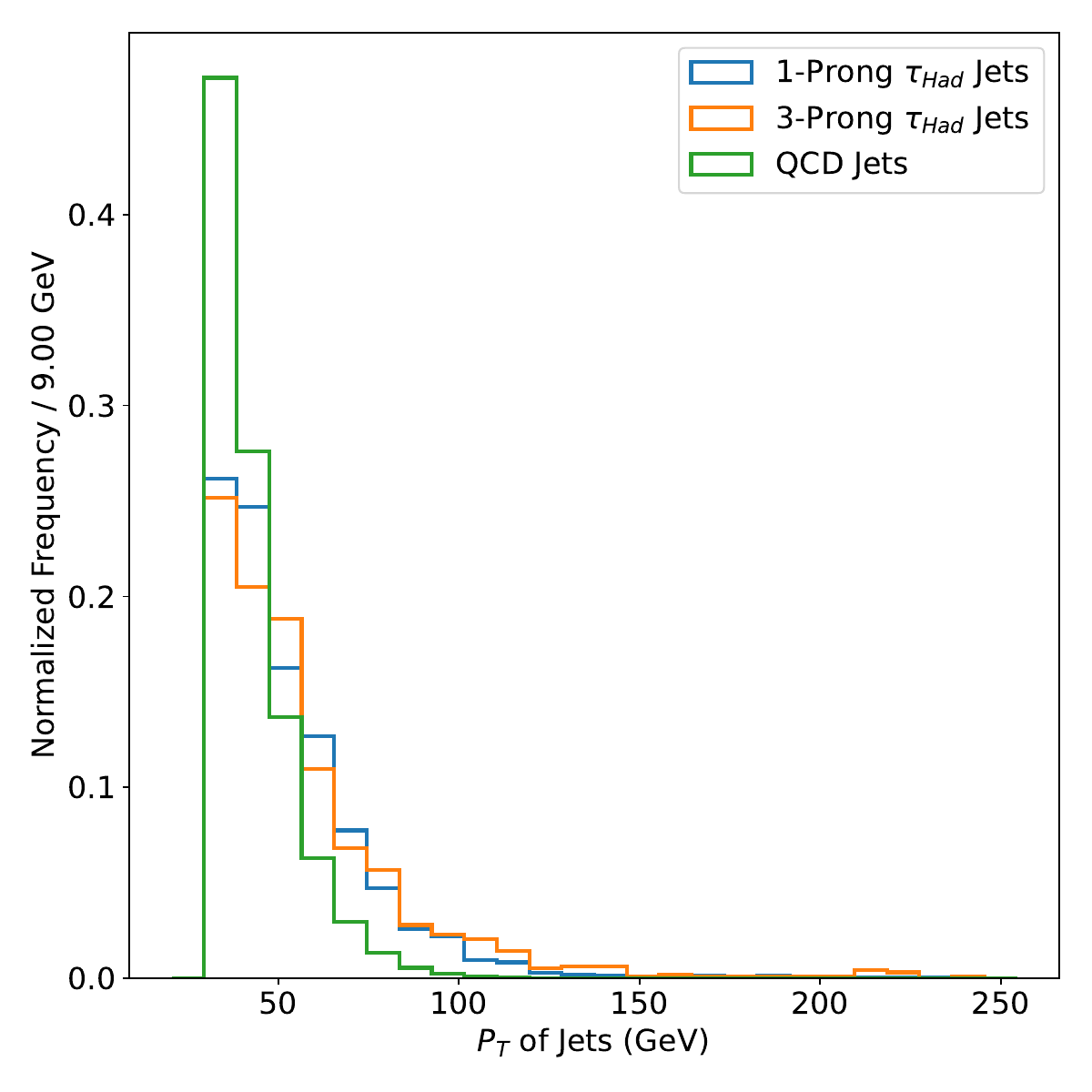}
    \includegraphics[width=0.31\textwidth]{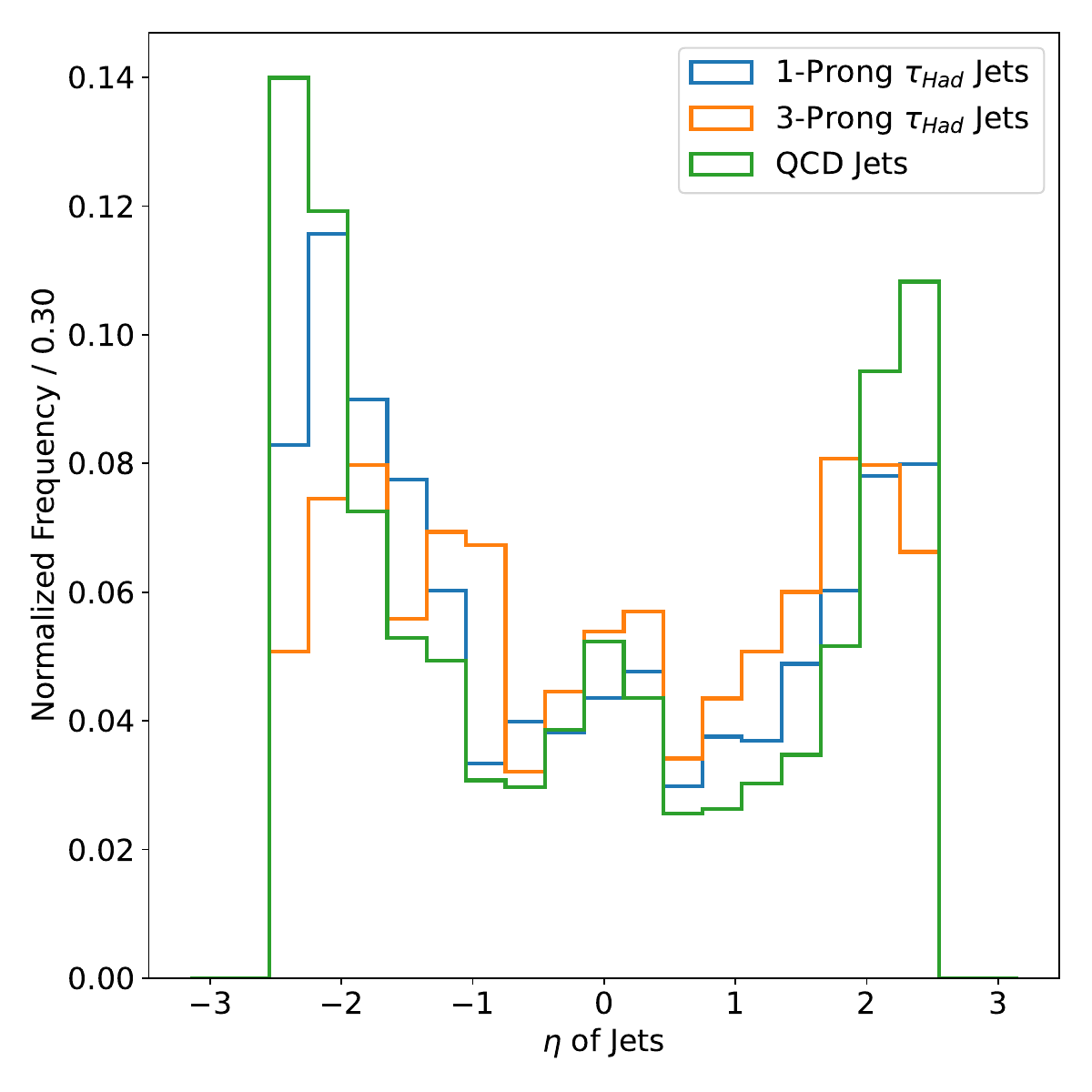}
    \includegraphics[width=0.31\textwidth]{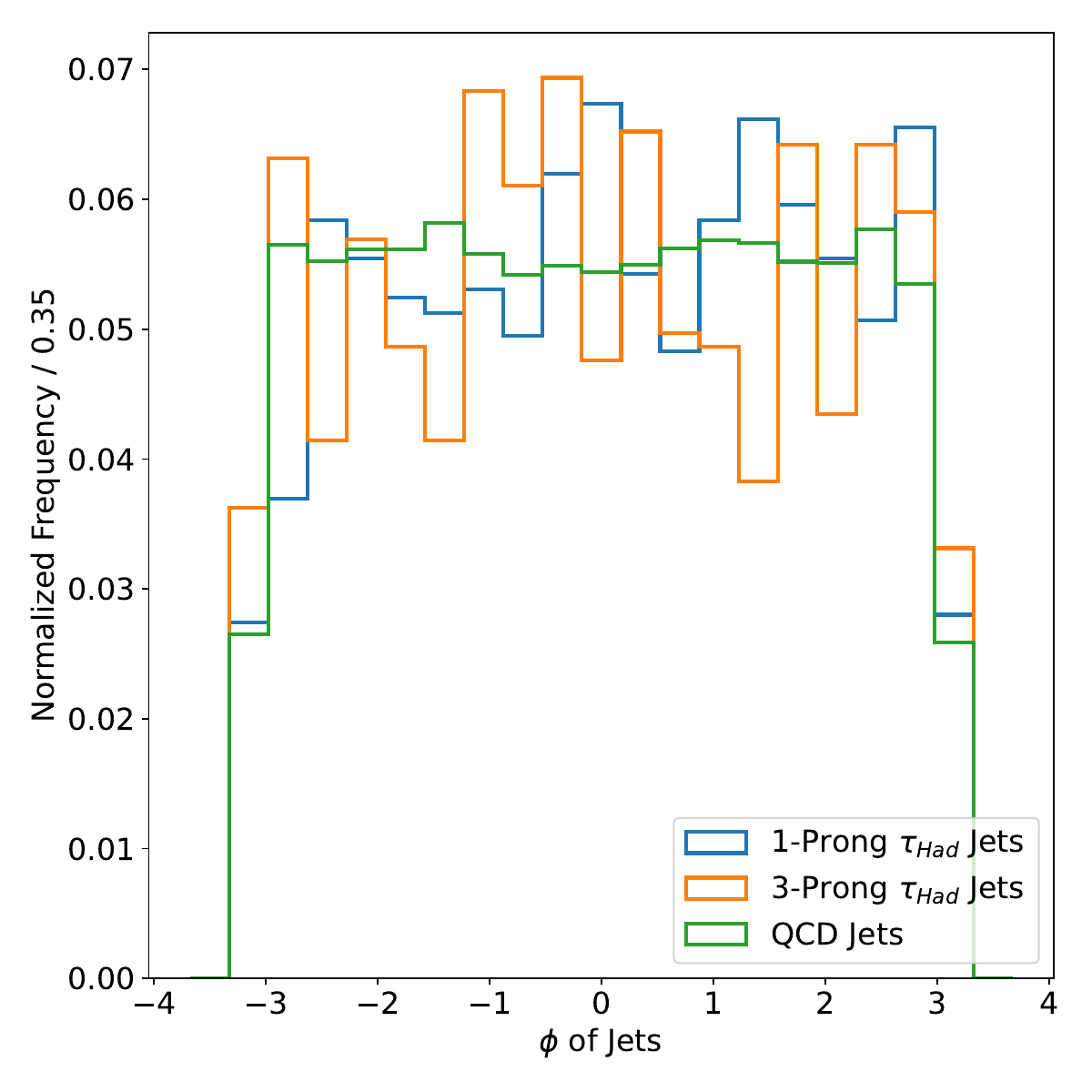}
    \caption{Comparison of the jet \pt, $\eta$, and $\phi$ of 1-Prong \taujet jets, 3-Prong \taujet jets, and background (QCD jets). All distributions are normalized to the same area.}
    \label{fig:jet_pt}
\end{figure}

Inputs to the neural networks for the tau identification are the detector-level track parameters (\trackpt, \tracketa, \trackphi, $d_0$, $z_0$) and tower kinematic variables (\towerpt, \towereta, \towerphi). The $d_0$ and $z_0$ are the transverse and longitudinal impact parameters with respect to the collision point. Figure \ref{fig:pt_et} shows the kinematic variables of the tracks and towers. We applied a $\log$ transformation to the \trackpt and \towerpt and normalized all input features between -1 and 1. In addition, the jet kinematic variables, namely the jet \pt, $\eta$, and $\phi$, are also used. And their distributions are shown in Figure~\ref{fig:jet_pt}.

\begin{figure}[htb]
    \centering
    \includegraphics[width=0.49\textwidth]{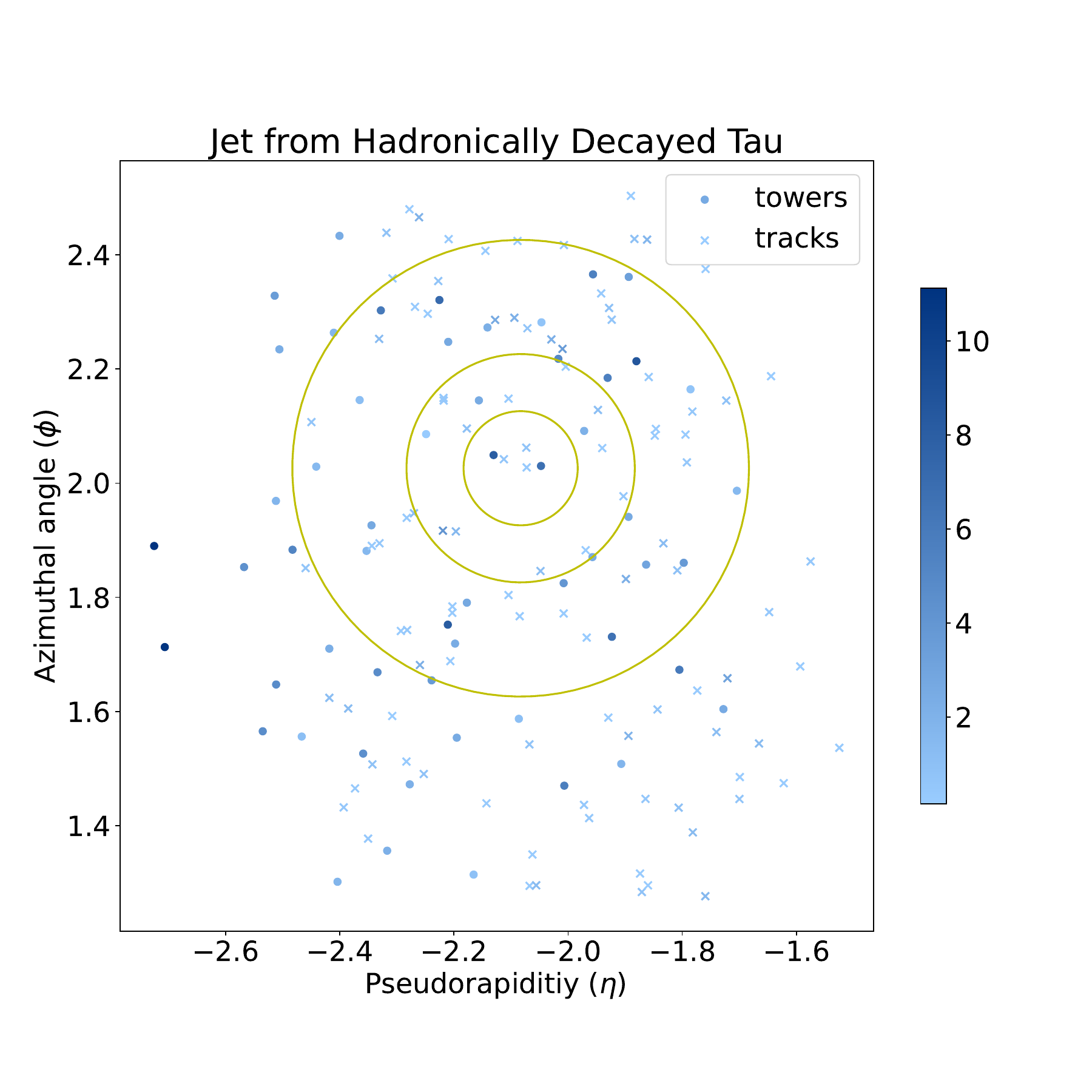}
    \includegraphics[width=0.49\textwidth]{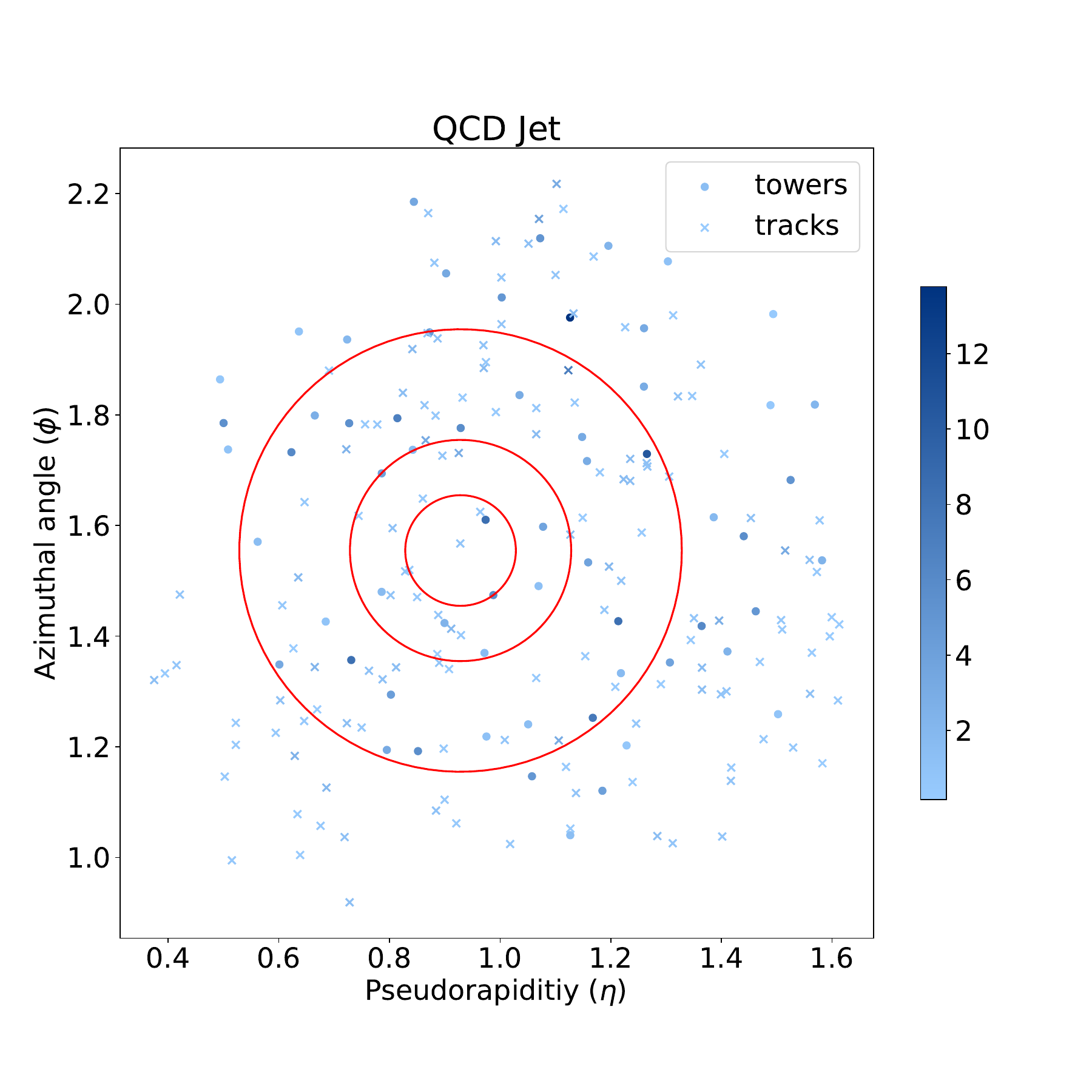}
    \caption{\taujet candidate jet vs. QCD jet, with the color of tracks and towers corresponding to \pt and $E_T$ respectively }
    \label{fig:jet_rep}
\end{figure}

Jets from the hadronically decayed tau leptons have geometrically different track and tower distributions than those from QCD. As shown in Figure \ref{fig:jet_rep}, tracks are more concentrated on the jet axis for \taujet jet than QCD jets. Therefore, we divide the regions encompassing the jet axis according to the distance $\Delta R$ away from the jet axis. We define the \textit{core region} as $\Delta R < 0.1$, the \textit{isolated (central) region} $0.1 \leq \Delta R < 0.2$, and the \textit{outer region} $0.2 \leq \Delta R < 0.4$. 

\begin{figure}[htb]
    \centering
    \includegraphics[width=0.31\textwidth]{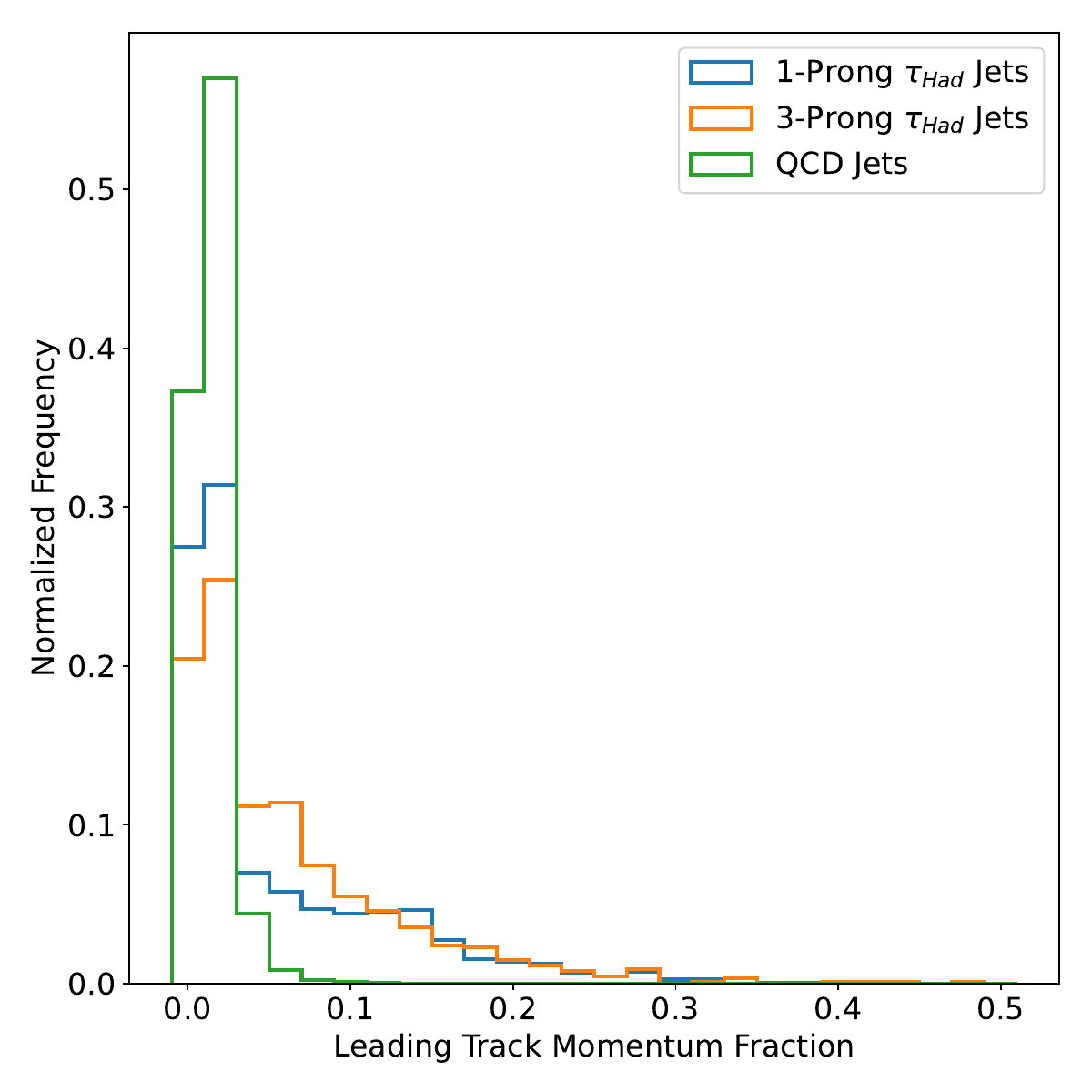}
    \includegraphics[width=0.31\textwidth]{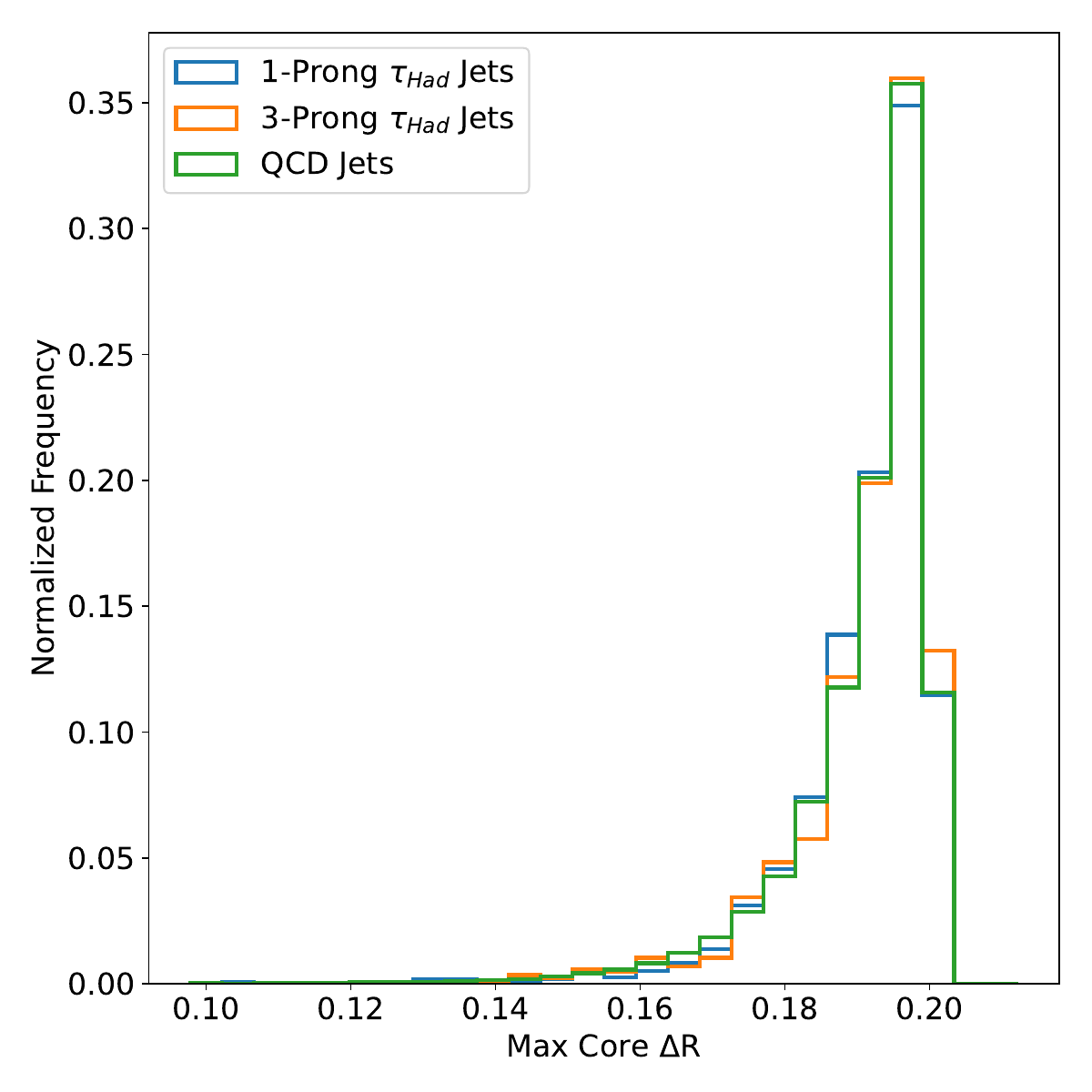} \\
    
    \includegraphics[width=0.31\textwidth]{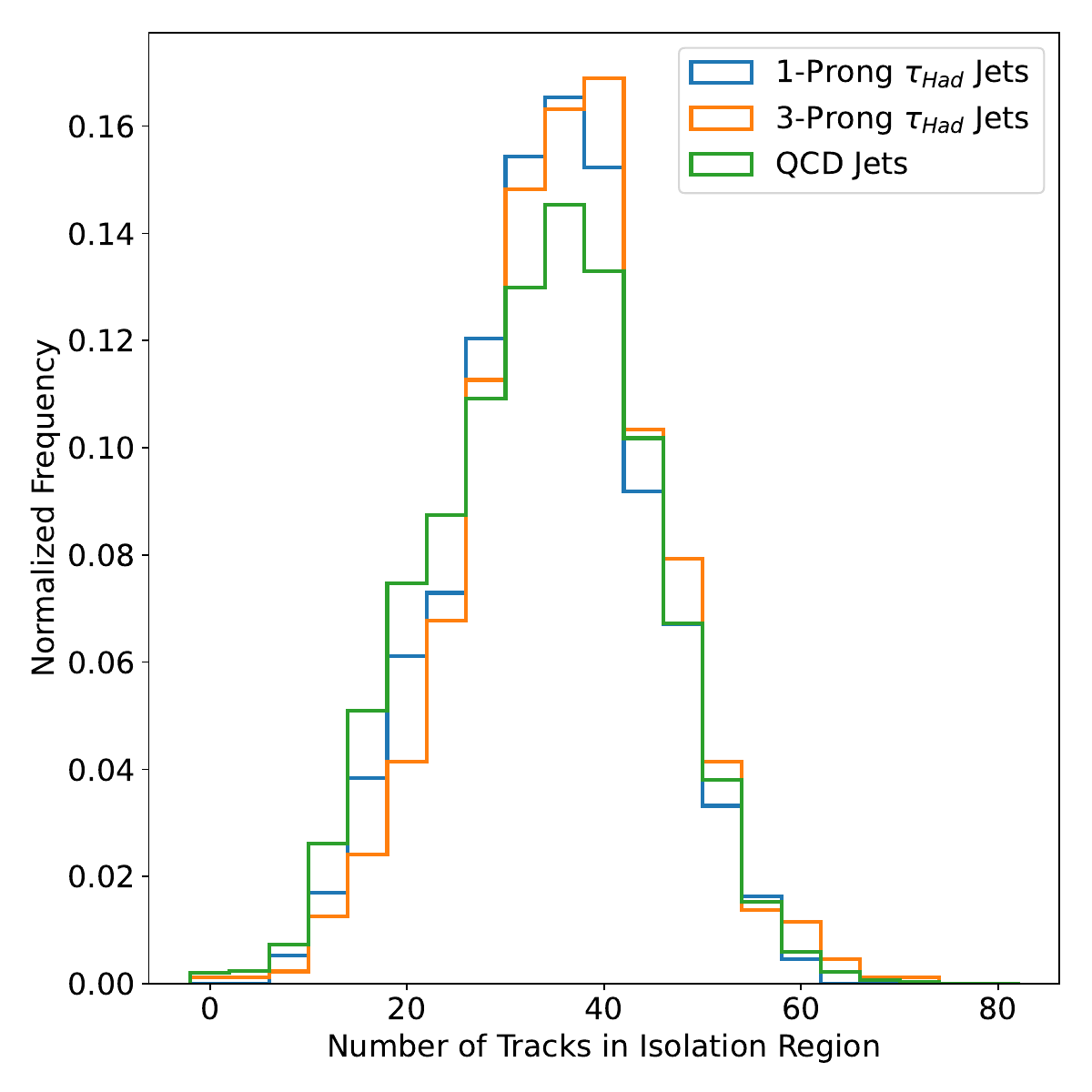}
    \includegraphics[width=0.31\textwidth]{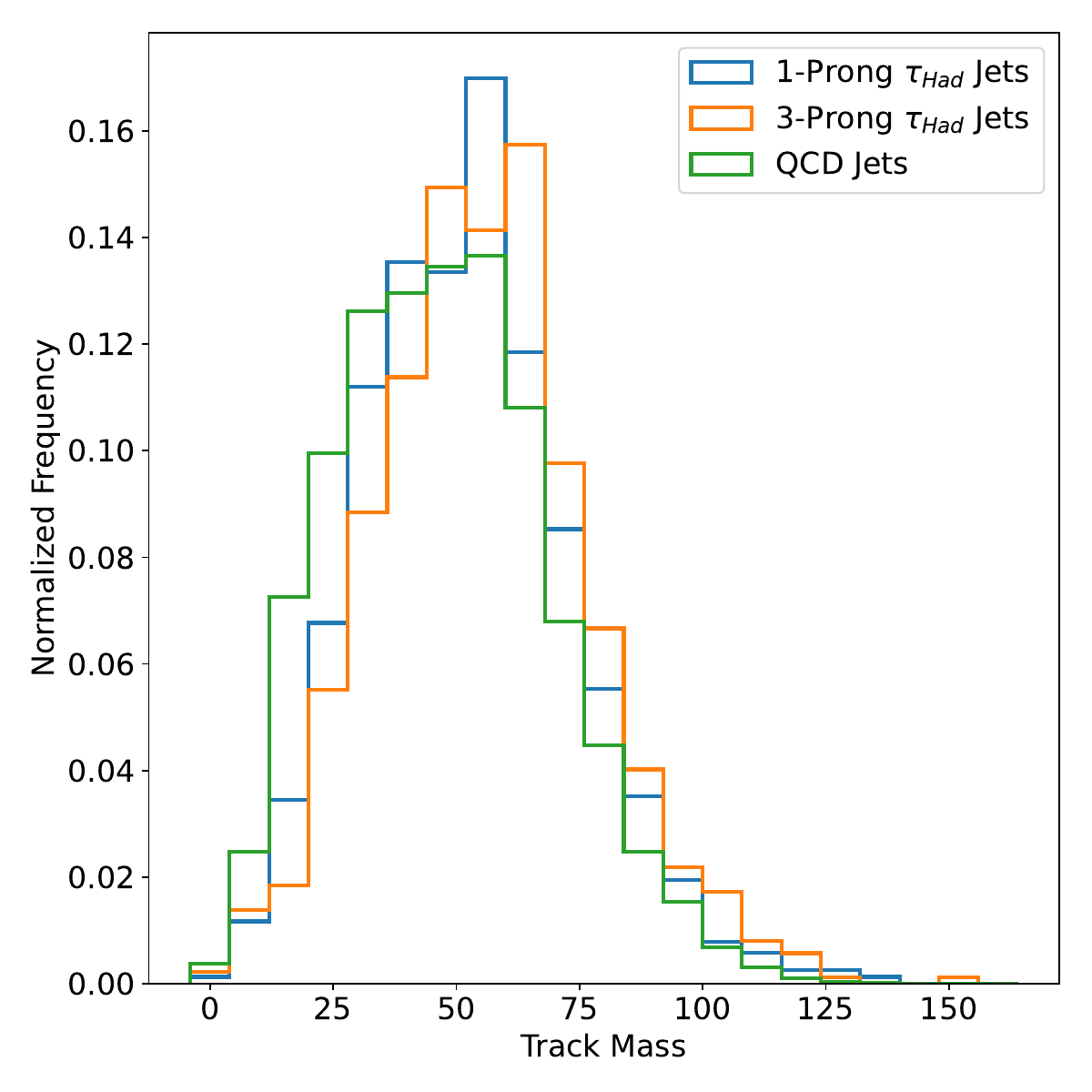}
    \includegraphics[width=0.31\textwidth]{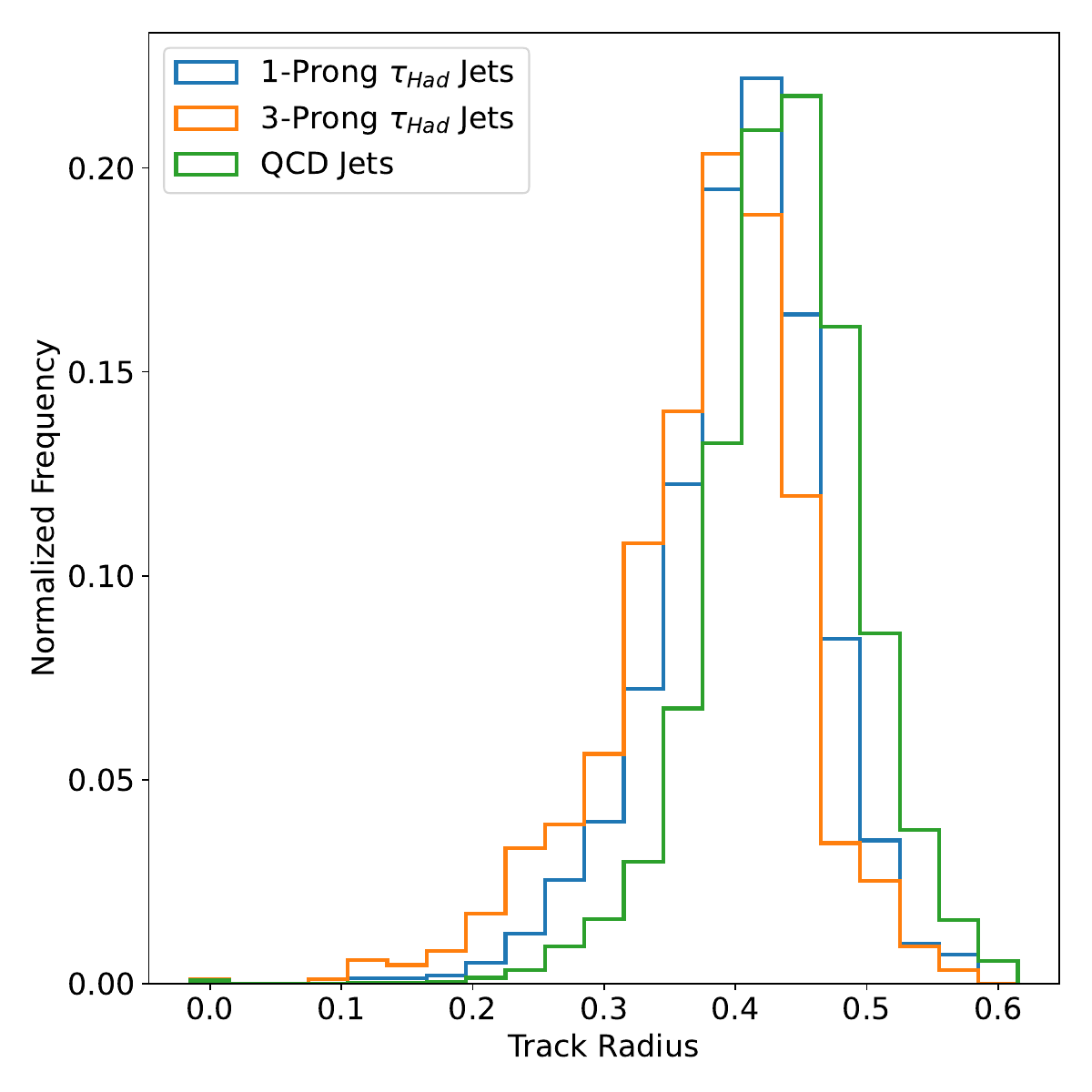}
    \caption{Comparison of distributions of high-level variables for 1-Prong \taujet jets, 3-Prong \taujet jets, and background (QCD jets). All distributions are normalized to the same area.}
    \label{fig:HLV}
\end{figure}

This further motivates us to calculate physics-inspired high-level variables defined in Table~\ref{tab:hlv_def}. Figure~\ref{fig:HLV} compares the distributions of the high-level variables among the 1-Prong \taujet, the 3-Prong \taujet, and the QCD jets.  
Although the 1-Prong \taujet and 3-Prong \taujet decay modes differ, their input features fed to the neural network are the same, as listed in Table~\ref{tab:graphs}. The same model can be applied to both decay modes; therefore, we trained one model for both decay modes to gain statistics.

\begin{table}[htb]
\caption{Definitions of high-level variables for hadronic \taujet identification.} \label{tab:hlv_def}
\centering
%\resizebox{\textwidth}{!}{
\begin{tabular}{|p{4cm}|p{1.5cm}|p{8cm}|}
\hline
    Name & Symbol & Definition \\ \hline
    Leading Track Momentum Fraction & \fracltrack & Transverse momentum of highest \pt track associated with jet divided by the sum of transverse energy found in the core region of the jet  \\ \hline
    Track Radius & \rtrack & \pt weighted distance of tracks associated with tau candidate from tau candidate direction \\ \hline
    Track Mass & \mtrack & Invariant mass calculated from the sum of the four-momenta of all tracks in the core and isolation region (assuming pion mass for each track) \\ \hline
    Number of Isolated Tracks & \nisotrack & Number of tracks in the isolated region of jet line line line line \\ \hline
    Maximum $\Delta R$  & \drtrack & Maximum distance between a core track associated with the tau candidate and the tau candidate direction \\ 
\hline 
      
\end{tabular}
%}
\end{table}

\section{Methods} \label{sec:method}

\subsection{Graph Neural Networks}
\label{subsec:gnn}
\begin{figure}[htb]
    \centering
    \includegraphics[width=0.9\textwidth]{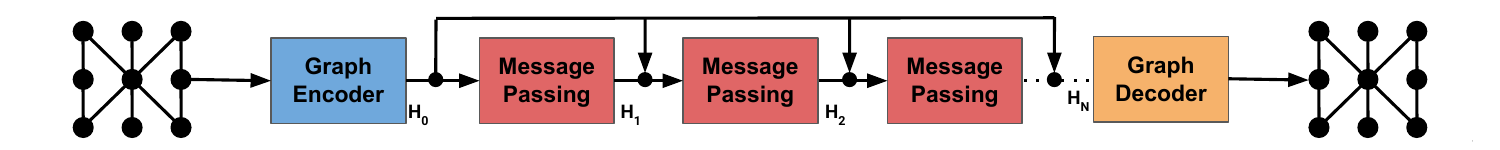}
    \caption{A general graph neural network architecture.}
    \label{fig:gnn}
\end{figure}

The GNN models we are exploring contain three discrete modules: Graph Encoder, Message Passing, and Graph Decoder, as pictured in Figure \ref{fig:gnn}. Each module contains three basic neural networks: node networks, edge networks, and global networks. These basic neural networks can be any network architecture. In our studies, they are two layers of Multi-Layer Perceptrons (MLPs), each with 64 units.

\textit{Graph Encoder} first updates input node features $v_i$ through a node network $\phi^v$:
$ v_i' = \phi^v (v_i) $, where $i$ loops over all nodes. Then, the graph encoder creates edge features using the concatenated features of the two connected nodes through an edge network: $e_k' = \phi^e (v_{r_k}, v_{s_k}) $, where $v_{r_k}$ and $v_{s_k}$ are the previously updated node features of the two nodes connected to the edge in question and $k$ loops over all edges. Finally, the graph-level attributes are updated independently through a graph-level network: $u' = \phi^{u}(u)$. 
After going through the graph encoder, the input graph becomes an attributed graph with latent node, edge, and graph-level features, denoted as $H_0$ in Figure~\ref{fig:gnn}. 

\textit{Message Passing} is designed to exchange information among the three-level graph attributes. Each message passing step updates the edge-, node-, and graph-level features sequentially. First, edge features at the $t$-th message passing step are updated: 
\[ e_k^{t} = \phi^e (e_k^0, e_k^{t-1}, v_{r_k}^{t-1}, v_{s_k}^{t-1}, u^{t-1})\]
where $e_k^0$ are the edge features after the encoder, $e_k^{t-1}$ are the edge features at the previous message passing step. Similarly,  $v_{r_k}^{t-1}$ and $v_{s_k}^{t-1}$ are the node features and $u^{t-1}$ are the global features. Then node features are updated. 
\[ v_i^t = \phi^v (v_i^0, v_i^{t-1}, \Bar{e}_i^{t}, u^{t-1}) \]
where $\Bar{e}_i^{t-1}$ is the aggregated edge features at the $t$-th message passing step. Finally, global features are updated: 
\[ u^t = \phi^u (u^0, u^{t-1}, \Bar{e}_i^{t}, \Bar{v}^{t}) \]
where $\Bar{v}^{t}$ is the aggregated node features at the $t$-th message passing step.

\textit{Graph Decoder} updates node, edge, and graph-level features independently. In the \taujet identification algorithm, Graph Decoder applies neural networks on the graph-level features to get a classification score.

\subsection{Homogeneous Graph Representations}

Each jet is represented as a fully-connected graph in which nodes are tracks or towers, and edges are connections between every two nodes. We explore different graph attribute assignments. The baseline is to assign node features with only the track or tower kinematic variables, denoted as $G_1$. Then we add the \jetpt to all nodes, $G_2$, and jet kinematics to graph-level attributes, $G_3$. Instead of using the absolute angular variables ($\eta$ and $\phi$) of tracks and towers, we use the relative angular variables to the track/tower nodes: \trackdeta and \trackdphi, leading to configurations of $G_4$, $G_5$, and $G_6$. Physics-inspired edge features, as used in Ref~\cite{Qu:2022mxj}, were added to the graphs as inputs. But no improvement was observed. Hence, no edge features are used in the input graphs.

Table~\ref{tab:graphs} summarizes different graph attribute assignments explored. To compare the performance of each graph representation, we use the same homogeneous GNN model, of which the detail is in Section~\ref{subsec:gnn}. Note that tower node features are padded with zeros to make the graphs homogeneous. The padding is not needed for heterogeneous GNNs.

\begin{table}[htb]
    \centering
    \caption{Different graph representations for jets. All graphs are full-connected graphs with different node and graph-level attributes. Track node features and tower node features are harmonized by padding the tower node features with zeros. $G_1$, $G_2$, and $G_4$ graph types do not have global (i.e. graph-level) attributes. See Table \ref{tab:hlv_def} for high-level variable definitions.}
    \label{tab:graphs}
    \resizebox{\textwidth}{!}{
    \begin{tabular}{|c|c|c|}
        \hline
         & $G_1$ & $G_2$  \\ \hline
        Track nodes & \trackpt, \tracketa, \trackphi, \dzero, \zzero & \trackpt, \tracketa, \trackphi, \dzero, \zzero, \textcolor{blue}{\jetpt} \\
        Tower nodes & \towerpt, \towereta, \towerphi, 0, 0           & \towerpt, \towereta, \towerphi, 0, 0, \textcolor{blue}{\jetpt} \\
        Global & None  &  None  \\ \hline 
        
         & $G_3$ & $G_4$ \\ \hline
        Track nodes & \trackpt, \tracketa, \trackphi, \dzero, \zzero, \jetpt & \trackpt, \textcolor{blue}{\trackdeta, \trackdphi,} \dzero, \zzero, \jetpt \\
        Tower nodes & \towerpt, \towereta, \towerphi, 0, 0 , \jetpt           & \towerpt, \textcolor{blue}{\towerdeta, \towerdphi}, 0, 0, \jetpt \\
        Global & \textcolor{blue}{\jetpt, \jeteta, \jetphi}  &  None  \\ \hline

        & $G_5$ & $G_6$  \\ \hline
        Track nodes & \trackpt, \textcolor{blue}{\trackdeta, \trackdphi,} \dzero, \zzero, \jetpt & \trackpt, \textcolor{blue}{\trackdeta, \trackdphi,} \dzero, \zzero, \jetpt \\
        Tower nodes & \towerpt, \textcolor{blue}{\towerdeta, \towerdphi}, 0, 0, \jetpt           & \towerpt, \textcolor{blue}{\towerdeta, \towerdphi}, 0, 0, \jetpt  \\
        Global & \textcolor{blue}{\jetpt, \jeteta, \jetphi}  &  \textcolor{blue}{\jetpt, \jeteta, \jetphi}, High-level Variables  \\ 

        \hline

    \end{tabular}
    }
\end{table}

\subsection{Heterogeneous Representations}
The differences between homogeneous GNNs and heterogeneous GNNs lie in whether the basic neural networks treat different node and edge types differently. For homogeneous GNNs, all nodes and edges are updated by the same neural networks. However, for heterogeneous GNNs, different neural networks are applied to different node types and edge types to perform encoding, message passing, or decoding. In our study, we mainly explore three different heterogeneous graph encoders: the Heterogeneous Node Encoder, the Heterogeneous Node and Edge Encoder, and the Recurrent Encoder.

The homogeneous GNN model uses the \homoenc{}, which treats all tracks and towers as the same type of objects and applies the same basic MLPs to each. The \hetnode{} treats the nodes as two types of objects, namely the tracks and towers, and applies two separate neural networks to each. The updated nodes are then passed into the same aggregation functions and edge networks. The \hetedge{} utilizes three distinct edge update functions corresponding to the three different edges in the graph, in addition to the two separate node update functions in the \hetnode{}. An illustration of this architecture is shown in Figure \ref{fig:het_enc}.

\begin{figure}[htb]
    \centering
    \includegraphics[width=0.9\textwidth]{Het_Enc_Illustration.pdf}
    \caption{Architecture of the \hetedge{}.}
    \label{fig:het_enc}
\end{figure}

The \recrenc{} is inspired from the Recurrent Neural Network architecture \cite{ATLAS:2019uhp}. Instead of a permutation-invariant encoding, this model encodes the tracks and towers separately as sequences, using two MLP embedding layers followed by two Long Short-Term Memory (LSTM) layers \cite{ML:LSTM}, as illustrated in Figure \ref{fig:rnn_enc}. All of the layers use 32 hidden units.
The sequential encodings require a fixed-length input, and hence we only use the first ten tracks and six towers inside the jets, ordered descendingly by their $P_T$ and $E_T$ values. In addition, the edges are not used in this encoding, and the global attributes are updated only based on the node and graph-level attributes. Hence, we do not use a \textit{Message Passing} module and directly pass the attributed graphs into a \textit{Graph Decoder} module.
\begin{figure}[htb]
    \centering
    \includegraphics[width=0.9\textwidth]{RNN_Enc_Illustration.pdf}
    \caption{Architecture of the \recrenc{}.}
    \label{fig:rnn_enc}
\end{figure}

In order to study the importance of the sequential relation used in the \recrenc{}, we also explore the \attnenc{} with and without Positional Encodings. The \attnenc{} has the same architecture as the \recrenc{}, except that the two LSTM layers are replaced with two Multi-Head Attention layers \cite{ML:Attention} with eight heads and embedding dimension of 32 units, where the first layer is a self-attention layer among the corresponding nodes, and the second layer conducts attention from the embedded graph-level attributes to the nodes. We then evaluate the importance of the sequential relation by adding Positional Encodings to the inputs before passing them into the MLPs. Table~\ref{tab:gnns} summarizes those GNN architectures.

\begin{table}[htb]
    \centering
    \resizebox{\textwidth}{!}{
    \begin{tabular}{|l|c|c|c|c|c|}
    \hline
        Models & Parameters & Input Graph & Message Passing Steps & AUC & Rejection at 75\% Efficiency \\ \hline
         \homoenc{} & 112,961 & $G_5$ & 1 & 0.9851 & 239.7 \\
         \hetnode{} & 130,049 & $G_5$ & 1 & 0.9864 & 339.6 \\
         \hetedge{} & 154,689 & $G_5$ & 1 & 0.9886 & 405.2 \\
         \recrenc{} & 62,801  & $G_6$ & 0 & 0.9932 & 4616.7 \\
         \attnenc{} with Positional Encoding & 154,329 & $G_6$ & 0 & 0.9926 & 1897.3 \\
    \hline
    \end{tabular}
    }
    \caption{Summary of different GNN architectures.}
    \label{tab:gnns}
\end{table}

\section{Results} \label{sec:results}

We evaluate the performance of different GNN models by examining the number of background QCD jets rejected at different signal \taujet selection efficiencies. Four working points at 45\%, 60\%, 75\%, and 95\% signal selection efficiencies are highlighted in the plots as points.

\subsection{Homogeneous Graph Representations}

\begin{figure}[htb]
    \centering
    \includegraphics[width=0.9\textwidth]{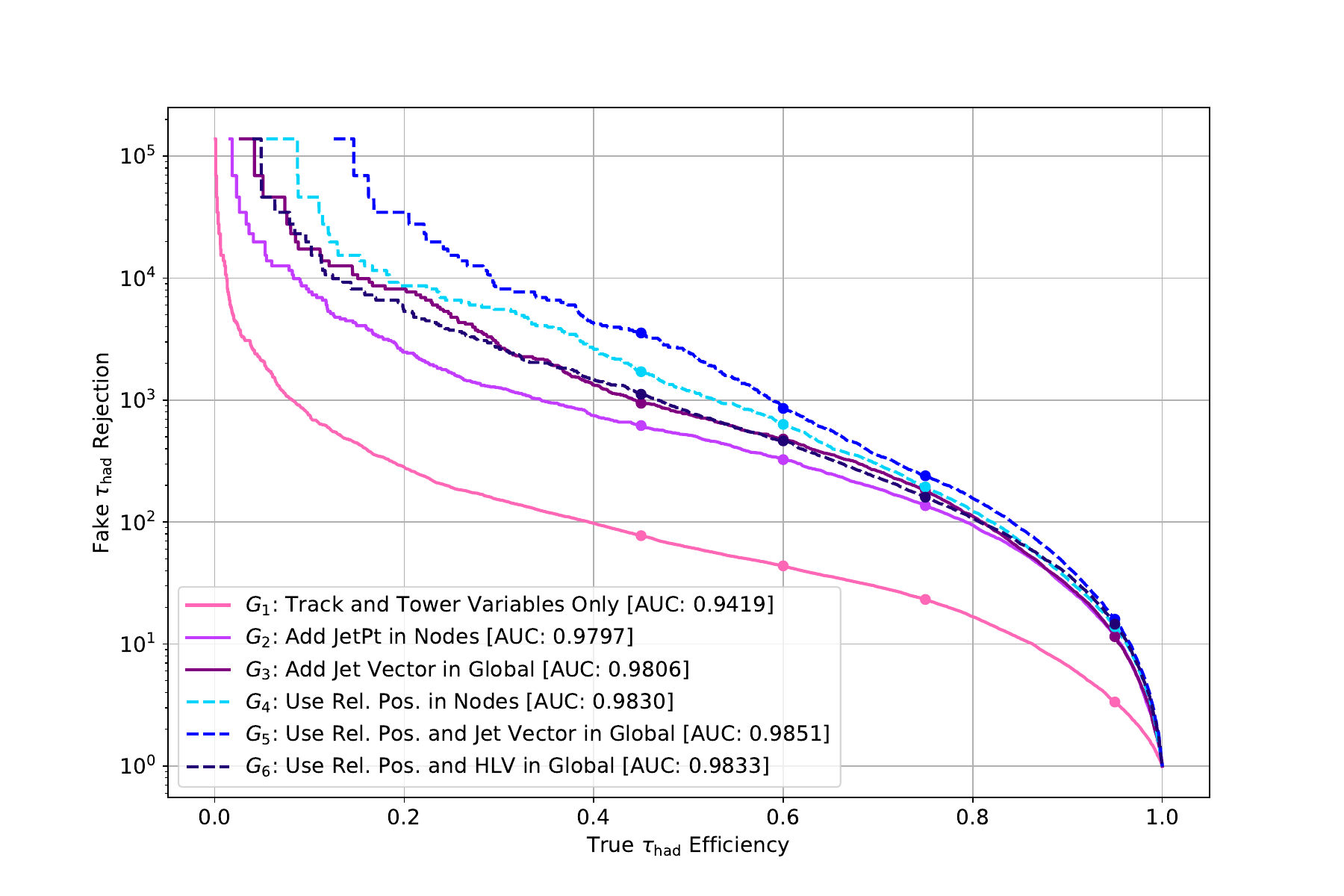}
    \caption{The number of background QCD jets rejected vs. signal \taujet selection efficiency for fully-connected, homogeneous graph network with different graph attributes. Curves with solid lines are graphs constructed with absolute track and tower positions as node inputs ($G_1$ through $G_3$), while curves with dashed lines are for graphs with relative positions ($G_4$ through $G_6$). Inputs for each configuration are listed in Table~\ref{tab:graphs}.}
    
    \label{fig:graph_rep}
\end{figure}

The performance of fully-connected, homogeneous graph networks acting on different graph representations is shown in Figure~\ref{fig:graph_rep}. 

By comparing the solid curves ($G_1$ through $G_3$), i.e., the models with absolute positions, we can see that the models gain a noticeable improvement when the jet-level variables are added as inputs, partially due to the separation power of the \jetpt. Although the \jeteta and \jetphi cannot separate the \taujet jet from the QCD jets, changing the $\eta$ and $\phi$ of the jet constituents to values relative to \jeteta and \jetphi results in better performance as concluded by comparing $G_2$ with $G_4$ and $G_3$ with $G_5$. The higher rejection power of $G_4$ at all working points indicates that using relative positions improves the model performance. By comparing the $G_4$ with the $G_5$, we can see that the model gains further improvement when jet-level variables are added as graph-level features even though the jet positions are already encoded in the nodes as relative positions.
Therefore, the jet-level variables deemed essential for the GNN models. However, it is worth noting that adding the derived high-level variables did not improve the GNN performance.

\subsection{Heterogeneous Encoders}

\begin{figure}[htb]
    \centering
    \includegraphics[width=0.9\textwidth]{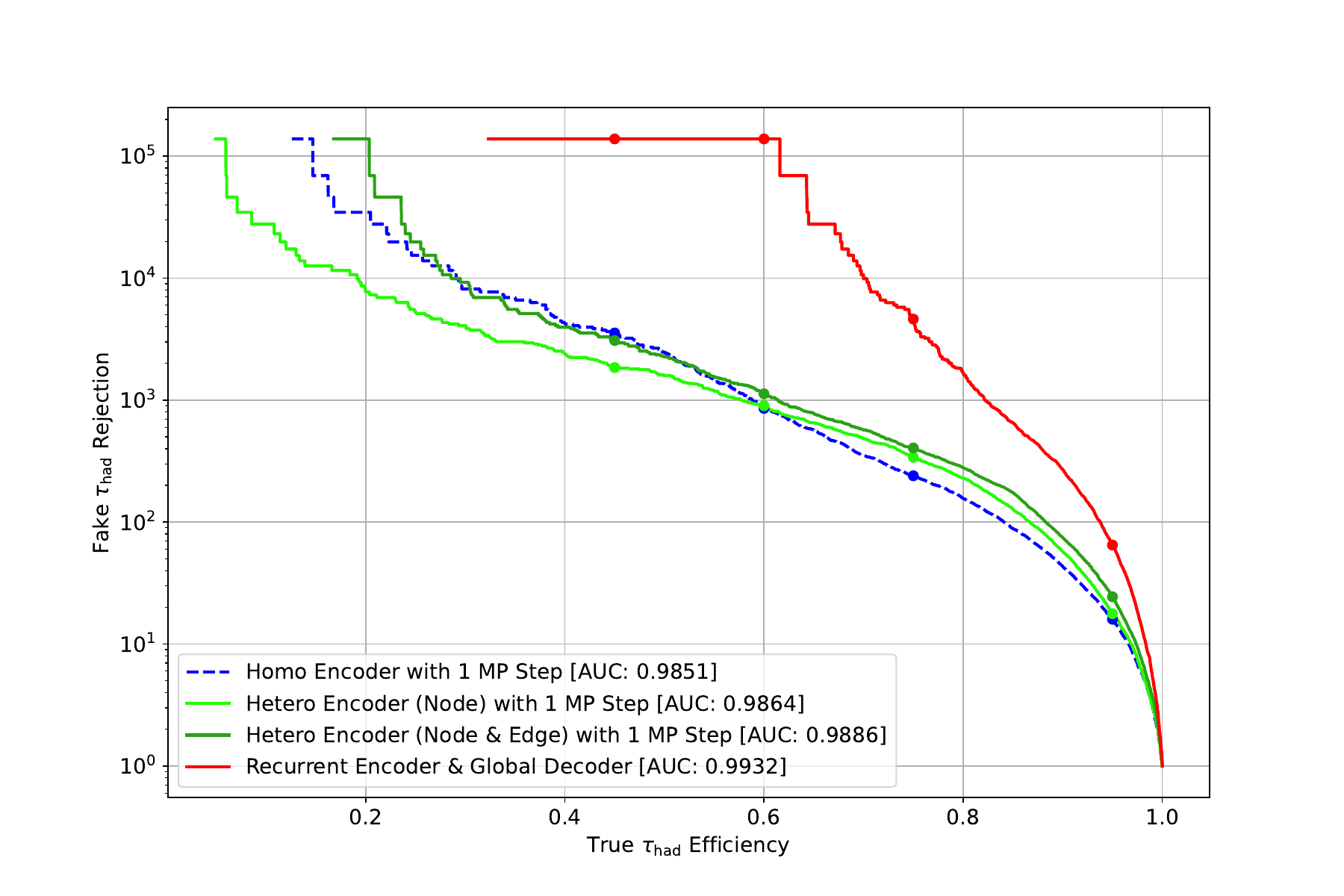}
    \caption{Number of background QCD jets rejected vs signal \taujet selection efficiency for homogeneous Graph Neural Network and heterogeneous Graph Neural Network. The dashed blue curve represents the homogeneous model presented as $G_5$ in Figure \ref{fig:graph_rep}. The solid curves represent models with heterogeneous encoders.}
    \label{fig:gnn_enc}
\end{figure}

All models use the input graph representation $G_5$ specified in Table \ref{tab:graphs}, except for the \recrenc{}, which uses $G_6$ and only the first 10 tracks and 6 towers as node inputs. The models with \textit{Message Passing} modules all use one message-passing step in this section. The performance of models with different encoders is presented in Figure \ref{fig:gnn_enc}.

It can be seen that the \hetnode{} rejects more background jets for higher efficiencies than the \homoenc{} but rejects fewer background jets for lower efficiencies. Specifically, the \hetnode{} rejects 100 more background jets (41.7\%) than the \homoenc{} at 75\% efficiency while rejecting 1705 fewer background jets (48.0\%) at 45\% efficiency. Compared to the \hetnode{}, the \hetedge{} rejects more background jets for all efficiencies, indicating that handling different types of edges using different MLPs improves the performance across all signal efficiency. It can also be seen that the \hetedge{} has a better rejection power than the \homoenc{} for higher efficiencies and similar rejection power for lower efficiencies, which yields a better AUC value and hence an indication of better overall performance. Noticeably, the \recrenc{} has significantly better rejection power than all other models, and potential explanations of this behavior are discussed in the following sections.

In addition, to check if the improvement of the heterogeneous encoder is a result of more trainable parameters, we also trained a \homoenc{} model with 178,241 parameters, comparable to the size of \hetedge{}. No significant improvement of the \homoenc{} is observed; the new AUC value is 0.9843, close to the original value and lower than that of \hetedge{}.

We also compares our results with the Particle Flow Network (PFN) model, a performant deep learning model for tagging top quarks using particle-level information~\cite{PFN,Kasieczka:2019dbj}. Although we use detector-level tracking and cluster information whilst the top quark taggers use particle-level hadron information, we take the PFN model and adjust the model size to be comparable with the \recrenc{} model. The comparison is shown in Table~\ref{tab:pfn}. Compared to the PFN model, the \recrenc{} provides more than a factor of two rejection powers for a signal efficiency of 75\%. 

\begin{table}[htb]
    \centering
    \resizebox{\textwidth}{!}{
    \begin{tabular}{|c|c|c|c|}
    \hline
        Models & Parameters & AUC & Rejection at 75\% Efficiency \\ \hline
        Particle Flow Network & 63,442 & 0.9907 & 1775.6 \\
        \hetedge{} & 154,689 & 0.9886 & 405.2 \\
       \recrenc{} & 62,801  & 0.9932 & 4616.7 \\
    \hline
    \end{tabular}
    }
    \caption{Comparison with Particle Flow Network in Ref.~\cite{PFN}.}
    \label{tab:pfn}
\end{table}

\section{Discussions}
\label{sec:discuss}
\subsection{Impact of Fixed-Length Inputs}

Since the \recrenc{} only accepts fixed-length input, we explore the impact of fixed-length inputs for other encoders by setting a cutoff in the number of nodes used, i.e., for a given input graph, changing the input node features from all tracks and towers associated with the jet to only the first 10 tracks and 6 towers, the same as the \recrenc{}. Comparing the \hetedge{} model with $G_5$ graph configuration, we observed that the model with cutoff yields a slightly worse rejection power than the model with all nodes, for all working points, where the AUC value is decreased from $0.9886$ to $0.9876$, and the rejection at $75\%$ efficiency decreased from 448 to 373. The other models also exhibit similar behavior. It is worth noting that applying the cutoff reduces the number of input nodes, thereby reducing the computation time in training from 16 hours to 12 hours.

\subsection{Message Passing Steps}

Due to the sequential relation, the LSTM layer embeds the $t^\text{th}$ node in the sequence by following the equation
\begin{equation*}
    o_t = \sigma(W_{io} x_t + W_{ho}h_{t-1} + b_o),
\end{equation*}
where $o_t$ is the output for the $t^\text{th}$ node, $\sigma$ is the sigmoid function, $x_t$ is the input of the $t^\text{th}$ node, $h_{t-1}$ is the hidden state of the $(t-1)^\text{th}$ node, and $W_{io}, ~W_{ho}, ~b_o$ are matrices for trainable parameters \cite{ML:LSTMmath}. This relation indicates that the embedding of the $t^\text{th}$ node is obtained by aggregating the hidden embedding of all the previous $t-1$ nodes \textit{recursively}.

In comparison, the permutation-invariant encoders update the nodes $v_i$ in each message-passing step by
\begin{equation*}
    v_i' = \phi^v (v_i, u, \Bar{e_i}'),
\end{equation*}
where $v_i'$ is the updated node embedding, $\phi^v$ is the node update function, $v_i$ is the node input, $u$ is the graph-level feature, and $\Bar{e_i}'$ is the aggregated edge variable obtained by aggregating
\begin{equation*}
    e_k' = \phi^e(e_k, v_{r_k}, v_{s_k}, u),
\end{equation*}
where $e_k'$ is the updated edge feature obtained by applying the edge updating function on the \textit{original} edge embedding $e_k$, the sender and receiver nodes $v_{r_k}$ and $v_{s_k}$ connected to edge $k$, and the graph-level feature $u$ \cite{battaglia2018relational}. In this framework, if only one message-passing step is used, even though the graph is fully connected, each node only aggregates the messages generated from ``old'' neighbors. A larger number of message-passing steps is needed for a node to aggregate messages sent along a path with a length greater than one.

We conducted the study using nine message-passing steps to ensure that each node can be updated by aggregating the messages along a path of length ten to match the cutoff length of the \recrenc{}. The result shows that the model with nine message-passing steps outperforms the model with one message-passing step for all working points; the rejection at $75\%$ efficiency is increased from $405$ to $449$, although the AUC values remain unchanged. It is worth noting that the performance of the model with more message-passing steps is still worse than the \recrenc{}.

\subsection{Influence of the Sequential Relation}

One major difference between the \recrenc{} and other permutation-invariant encoders is the sequential encoding used by LSTM. We explore the influence of the sequential encoding by comparing the performance of the \attnenc{} with and without the Positional Encoding \cite{ML:Attention}.

We find that the \attnenc{} with Positional Encoding outperforms the model without Positional Encoding for all working points. The AUC value and rejection power at $75\%$ efficiency of the model with Position Encoding are $0.9926$ and $1987$, higher than the model without Position Encoding ($0.9921$ and $1799$, respectively). Therefore, the sequential relation is likely beneficial for identifying \taujet jets. However, since the difference in the performance is not significant, the sequential relation may not be the only factor that determines the good performance of the \recrenc{}.

In addition, we further explore the contribution of each node in the sequence by examining the learned attention scores. The score is obtained from the second Multi-Head Attention layer, where the attention is conducted when aggregating node features to global attributes. In other words, the attention score represents the contribution of each node feature to the aggregated global attributes, which are then used for classification. We here extracted the attention score from a sample of 10,000 \taujet and QCD jets to visualize the contribution of nodes.

\begin{figure}[htb]
    \centering
    \includegraphics[width=0.45\textwidth]{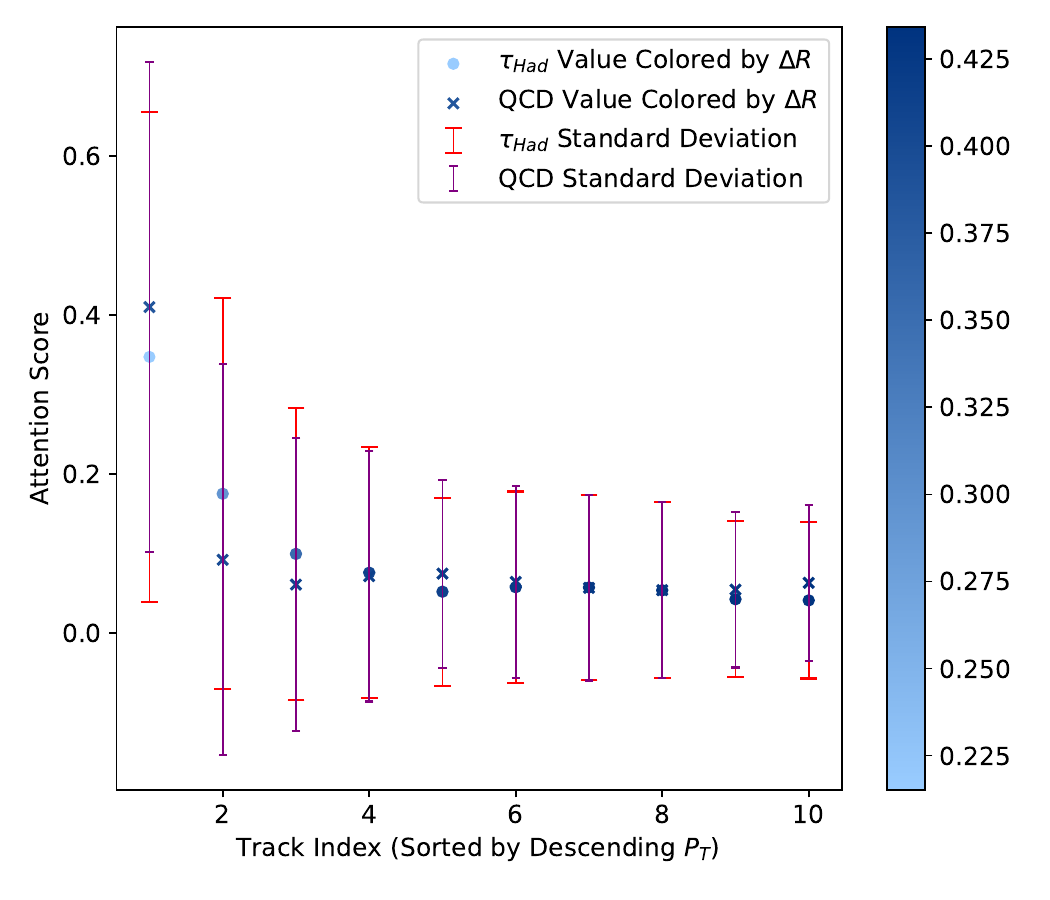}
    \includegraphics[width=0.45\textwidth]{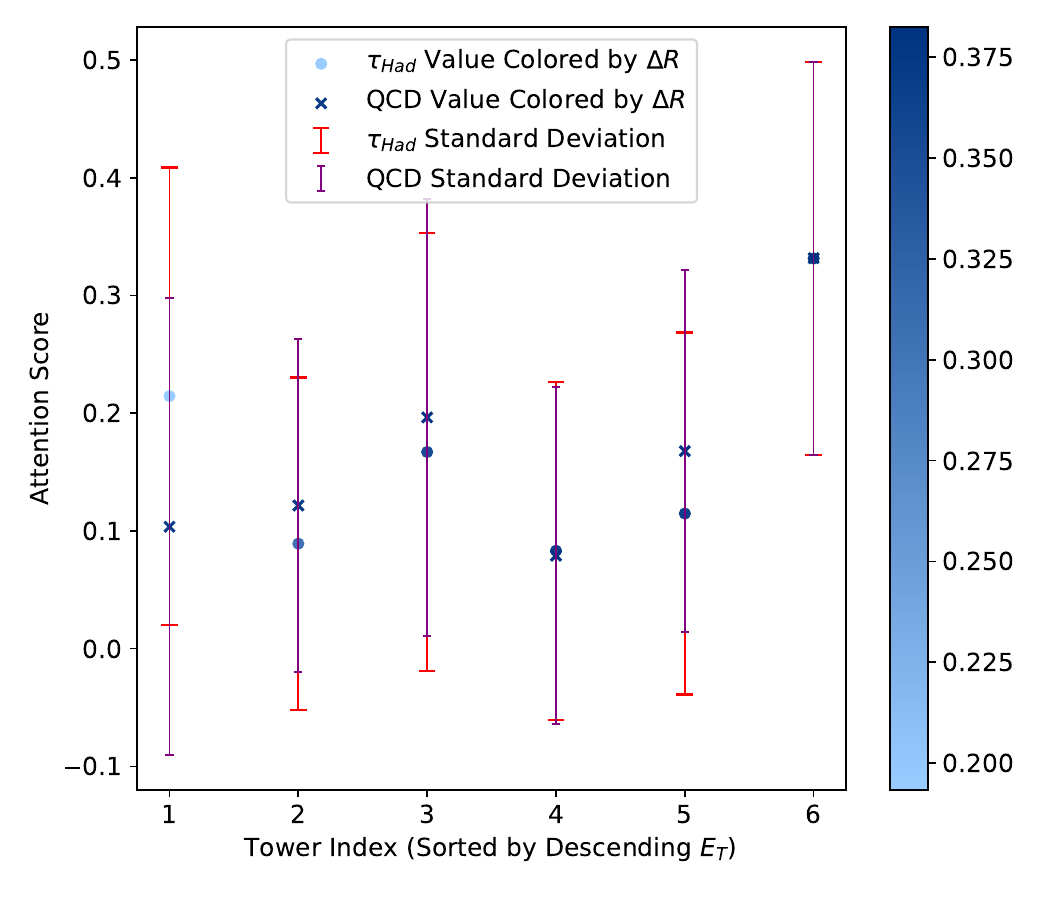}
    \includegraphics[width=0.45\textwidth]{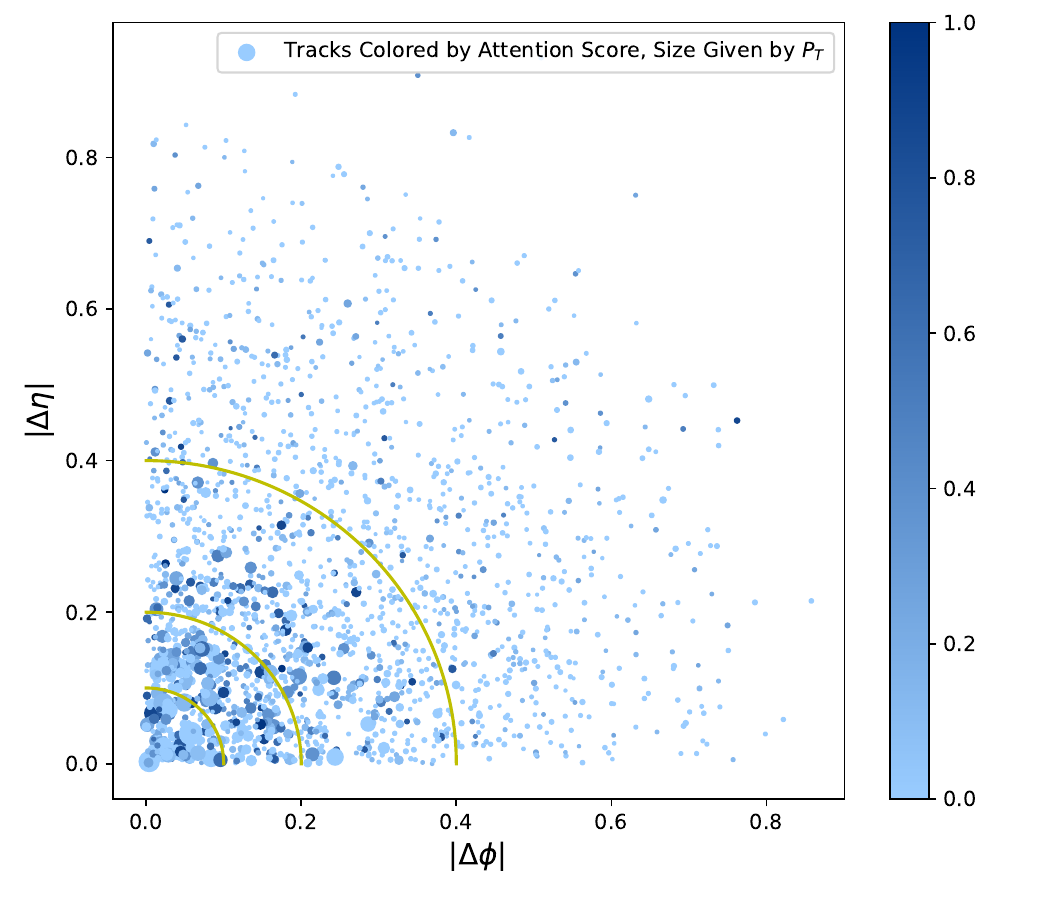}
    \includegraphics[width=0.45\textwidth]{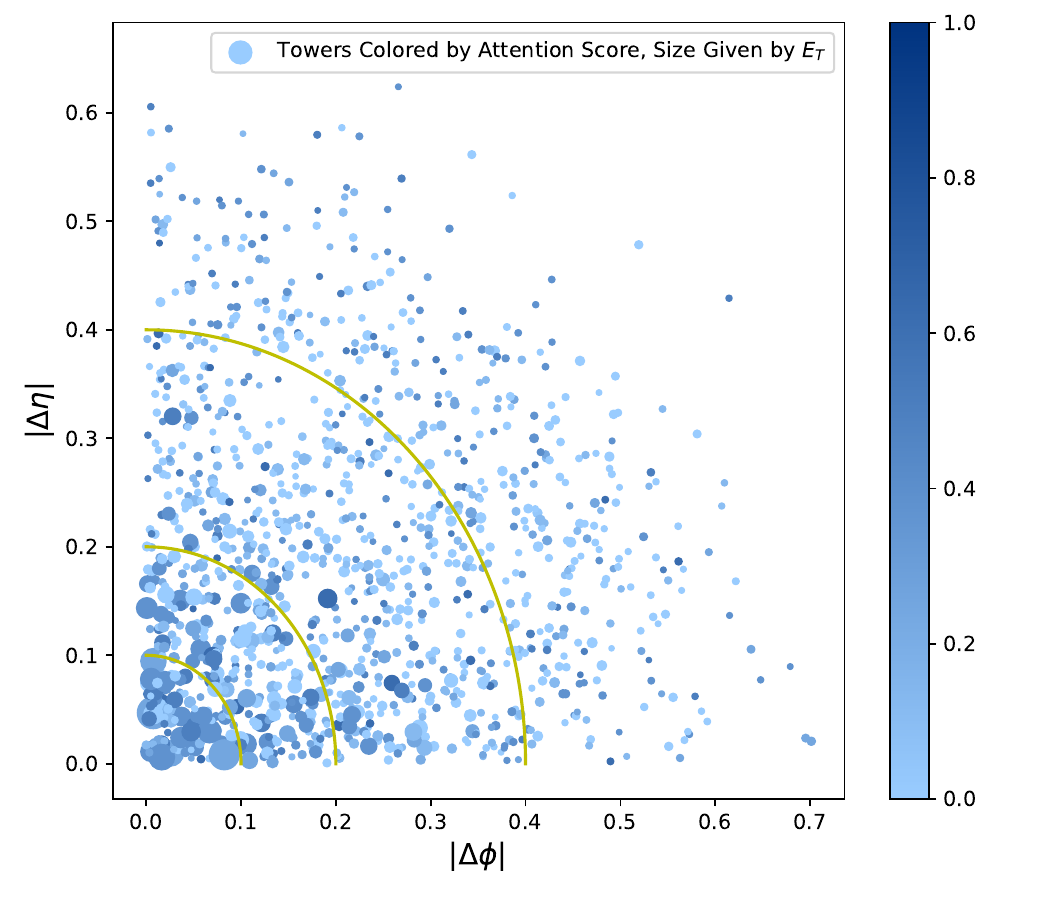}
    \caption{Top row: The mean attention scores of the tracks (left) and towers (right) for $10,000$ jets with standard deviations.  Bottom: The attention scores of tracks (left) and towers (right) for $200$ tau jets; the axes represent the absolute difference in coordinates from the jet axis. Each dot represents a track or tower whose colors represent the attention score, and sizes represent the corresponding $P_T$ or $E_T$ values.}
    
    \label{fig:attn_score}
\end{figure}

Figure~\ref{fig:attn_score} shows that the nodes with different \pt, at different sequence positions, contribute to the aggregated global attributes with different attention scores. On the one hand, the high \pt tracks are scored higher than those low \pt tracks, and the 2nd and 3rd highest \pt tracks are considered more critical for \taujet than QCD jets. However, the pattern is much less distinguishable for energy towers. And on the other hand, energy towers in the core regions are systematically scored higher than the central and outer regions, as shown in the bottom part of Figure~\ref{fig:attn_score}. However, the score distributions for tracks do not show a clear pattern in the $\eta$ and $\phi$ plane.

\subsection{Effects of Pileup}

\begin{figure}[htb]
    \centering
    \includegraphics[width=0.45\textwidth]{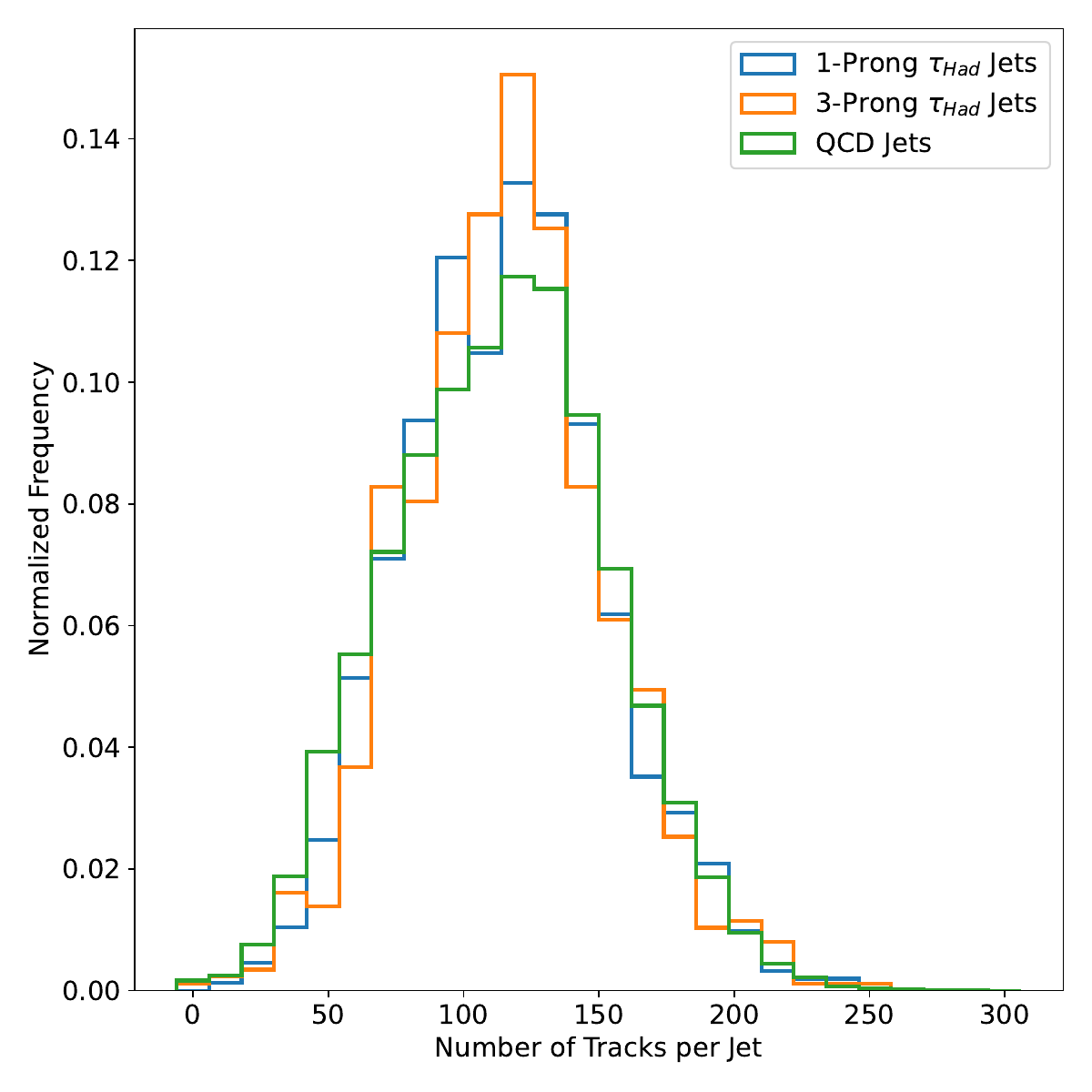}
    \includegraphics[width=0.45\textwidth]{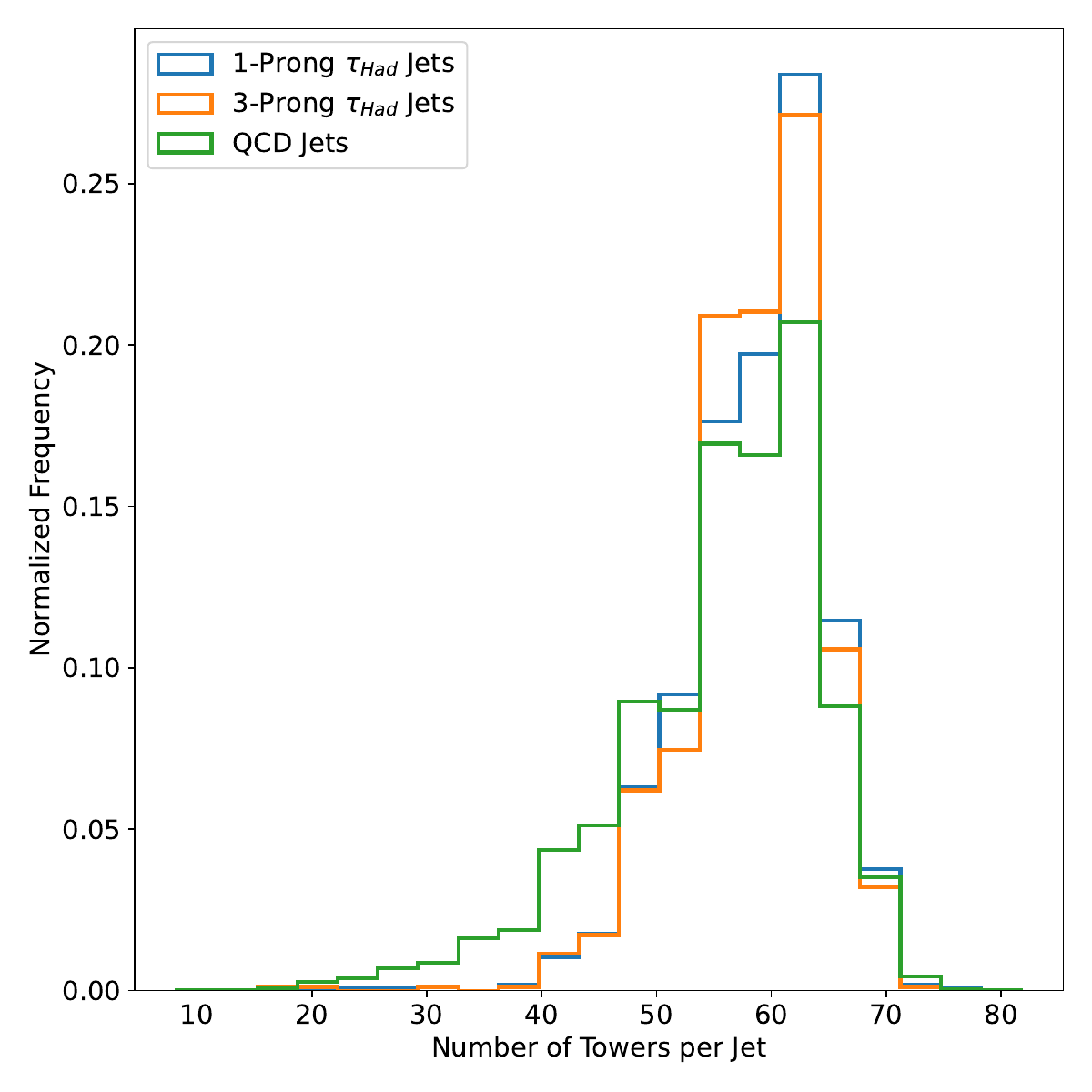}
    \includegraphics[width=0.45\textwidth]{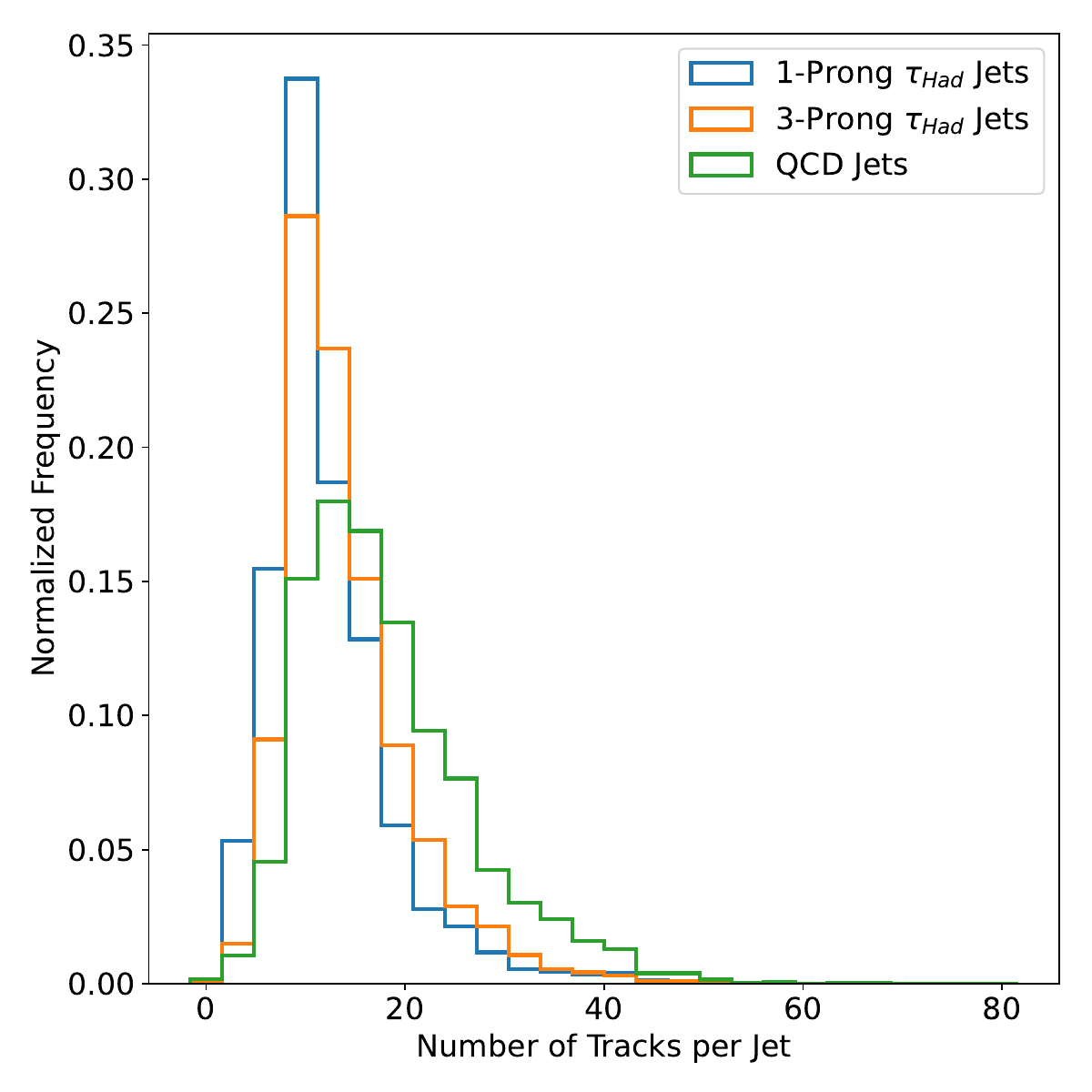}
    \includegraphics[width=0.45\textwidth]{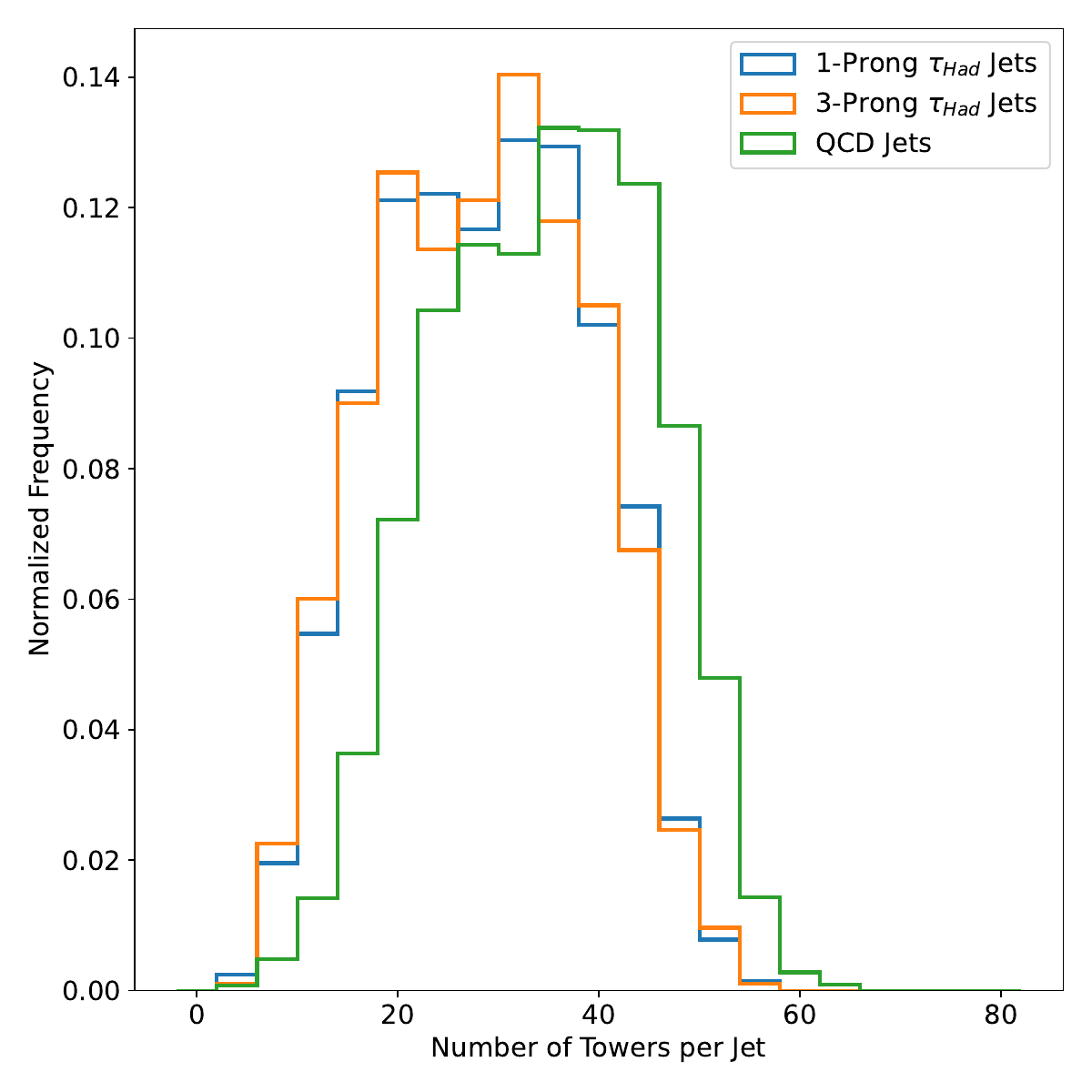}
    \caption{Comparison of the number of tracks and towers of 1-Prong \taujet jets, 3-Prong \taujet jets, and background (QCD jets). The top row represents the distributions for the $\mu=200$ dataset, and the bottom row represents the distributions for the $\mu=40$ dataset. All distributions are normalized to the same area.}
    \label{fig:num_trk_twr}
\end{figure}

In this section, we explore the effects of additional interactions resulting from pileup by examining models trained with a low-pileup dataset and applied to a high-pileup dataset and vice versa. The low-pileup dataset often leads to fewer tracks and towers inside the jets than the high-pileup dataset, as shown in Figure~\ref{fig:num_trk_twr}. 

The resulting AUC values for the \hetedge{} and \recrenc{} models are shown in Table~\ref{tab:pileup}.
Both models see a significant downgrade in the inference AUC values when the testing dataset differs from the training data, indicating poor generalizations of these models. The model trained with a high pileup dataset is better generalized than that trained with a low pileup dataset. The pileup dependence can be mitigated by training these models with a broader range of pileup conditions or with dedicated pileup suppression techniques, such as the Pileup Mitigation with Machine Learning~\cite{Komiske:2017ubm}.

\begin{table}[htb]
\caption{AUC values for models trained on a high pileup dataset and applied to a low pileup dataset, where $\mu$ denotes the average additional proton-proton collisions. The inferences are all conducted on the $\mu=200$ dataset.} \label{tab:pileup}
\centering
\resizebox{\textwidth}{!}{
\begin{tabular}{|c|c|c|c|c|}
    \hline
    Model & Training Dataset & Inference Dataset & AUC & Rejection at 75\% Efficiency \\ \hline

    Heterogeneous Node      & $\mu = 200$ & $\mu = 200$ & 0.9886 & 448.5 \\ 
        and Edge Encoder    & $\mu = 200$ & $\mu = 40$ &  \textbf{0.9804} & \textbf{107.3} \\ 
                            & \textbf{$\mu = 40$} & \textbf{$\mu = 200$} &\textbf{0.9614} & \textbf{32.8} \\ 
                            & \textbf{$\mu = 40$} & \textbf{$\mu = 40$} &0.9900 & 568.2 \\ 
                      
                                              \hline
    Recurrent & $\mu = 200$ & $\mu = 200$ & 0.9932 & 4616.7 \\ 
    Encoder   & \textbf{$\mu = 200$} & $\mu = 40$ & \textbf{0.9862} & \textbf{1033.3}\\ 
              & \textbf{$\mu = 40$} & \textbf{$\mu = 200$} & \textbf{0.9683} & \textbf{160.7}  \\
              & \textbf{$\mu = 40$} & \textbf{$\mu = 40$} &0.9928 & 487.1 \\ 
              \hline
\end{tabular}
}
\end{table}

%===================================================================
\section{Conclusions and Outlook} \label{sec:conclusions}
%===================================================================

In this study, we present the results of using a heterogeneous Graph Neural Network to identify \taujet jet against QCD jets for the HL-LHC. After examining various graph representations and heterogeneous encoders, we found that the jet-level information is essential for model's performance as adding the jet-level features to input graphs improves the model's separation power. While exploring the heterogeneity of the detector data and its integration into the Graph Neural Network, we found utilizing the \hetedge{} architecture results in increased QCD jet rejections in regions of high \taujet efficiency and comparable rejections in regions of lower \taujet efficiency, outperforming the \homoenc{} architecture overall. Additionally, we observed that sequential encoding outperforms permutation-invariant encodings because high \pt tracks and energy clusters in the core region are more important and need to be treated differently. 

\acknowledgments

This research used resources of the National Energy Research Scientific Computing Center (NERSC), a U.S. Department of Energy Office of Science User Facility operated under Contract No. DE-AC02-05CH11231.

\newpage
\bibliography{graph,main,tau,generators,HEPML}
\bibliographystyle{JHEP}

\clearpage

\appendix

\end{document}